\documentclass[usenatbib]{mn2e}
\usepackage{graphicx}
\usepackage{epsfig,psfrag}
\usepackage{natbib}
\title[HOD modeling for AGN]{The Halo Occupation Distribution of Active Galactic Nuclei} 
\author[Chatterjee et al. ]{\parbox{17cm}{
Suchetana Chatterjee$^{1}$, Colin DeGraf$^{2}$, Jonathan Richardson$^{1,3}$, Zheng Zheng$^{3,4}$, Daisuke Nagai$^{1,3}$, Tiziana Di Matteo$^{2}$}
\vspace{0.2cm}\\
$^{1}${Department of Astronomy, Yale University, New Haven, CT 06520 USA}\\
$^{2}${McWilliams Center for Cosmology, Carnegie Mellon University, Pittsburgh, PA 15213 USA}\\
$^{3}${Yale Center for Astronomy and Astrophysics, Department of Physics, Yale University, New Haven, CT 06520 USA}\\
$^{4}${Department of Physics and Astronomy, University of Utah, 115 South 1400 East, Salt Lake City, UT 84112}\\
}

\begin{document}
\maketitle

\begin{abstract}
Using a fully cosmological hydrodynamic simulation that self-consistently incorporates the growth and feedback of supermassive black holes and the physics of galaxy formation, we examine the effects of environmental factors (e.g., local gas density, black hole feedback) on the halo occupation distribution of low luminosity active galactic nuclei (AGN). We decompose the mean occupation function into central and satellite contribution and compute the conditional luminosity functions (CLF). The CLF of the central AGN follows a log-normal distribution with the mean increasing and scatter decreasing with increasing redshifts. We analyze the light curves of individual AGN and show that the peak luminosity of the AGN has a tighter correlation with halo mass compared to instantaneous luminosity. We also compute the CLF of satellite AGN at a given central AGN luminosity. We do not see any significant correlation between the number of satellites with the luminosity of the central AGN at a fixed halo mass. We also show that for a sample of AGN with luminosity above $10^{42}$ ergs/s the mean occupation function can be modeled as a softened step function for central AGN and a power law for the satellite population. The radial distribution of AGN inside halos follows a power law at all redshifts with a mean index of $-2.33\pm 0.08$. Incorporating the environmental dependence of supermassive black hole accretion and feedback, our formalism provides a theoretical tool for interpreting current and future measurements of AGN clustering.
\end{abstract}
 

\section{Introduction}
Through recent observations it has been shown that active galactic nuclei (AGN) play a significant role in the evolution of galaxies. The observed correlation between the mass of the central black hole and the velocity dispersion of the bulge of the host galaxy suggests a strong connection between galaxy evolution and black hole activity \citep[e.g.,][]{gebhardtetal00, m&f01, tremaineetal02, grahametal11}. Several theoretical models relating AGN activity or black hole growth to galaxy evolution have been proposed \citep[e.g.,][] {soltan82, s&r98, saluccietal99, k&h00, w&l03, marconietal04, cattaneoetal06, crotonetal06, dimatteoetal05, hopkinsetal06, lapietal06, shankaretal04}. Differentiating between these theoretical models requires several observational quantities. Measurements of the AGN luminosity function \citep[e.g.,][] {boyleetal00, fanetal01} and the number density of black hole hosts in the present universe can provide an estimate of the duty cycle of black holes. Alternatively, the black hole mass function \citep[e.g.,][]{shankaretal04, g&d07} measured at the current epoch can provide constraints on models of black hole growth. Measurement of AGN clustering provides a unique way to study the physical characteristics of AGN through the connection with their host dark matter halos \citep[e.g.,][]{croometal04, gillietal05, myersetal06, shenetal09, rossetal09}.

Theoretically clustering properties of AGN have been mostly studied with semi-analytic models using the halo model or the black hole continuity equation approach \citep[e.g.,][] {shankaretal10, bonolietal09, lidzetal06}. Although semi-analytic models have provided the formalism for interpreting quasar clustering, there are certain issues that these models cannot approach. In this formalism physical parameters (e.g.,``quasar duty cycle") are treated as free parameters that are constrained from clustering measurements \citep[e.g.,][]{mw01}. There also exist degeneracies between parameters in the models \citep[e.g.,][]{shankaretal10b}. Moreover in most of these models black holes are accreting at a fixed fraction of the Eddington rate, making accretion rate dependent on black hole mass only. In reality, parameters like accretion efficiency and the duty cycle are not fixed and depend strongly on environment (e.g., local gas density, mergers, feedback from AGN). Some recent semi-analytic studies started to consider more general cases of black hole accretion to investigate the cosmological co-evolution of black holes and the evolution of AGN luminosity functions \citep[e.g.,][]{marullietal08, marullietal09, malbonetal07, mencietal08}.  

Hydrodynamic simulations of galaxy formation with black hole growth can capture the environmental dependence of accretion. The growth of black holes is tied with the full dynamics of dark matter. Also with these simulations we can model the interplay between AGN and galaxy formation in the form of feedback and self-regulated growth \citep[e.g.,][]{w&l03}. Thus these simulations provide an excellent platform to study the co-evolution of AGN with large scale structures over cosmological time scales. Studies of AGN clustering using cosmological simulations have been carried out by \citet{thackeretal09} and \citet{degrafetal11a}. \citet{degrafetal11a} calculates the correlation function of black holes using a smoothed particle hydrodynamics (SPH) cosmological simulation that incorporates galaxy formation physics and self-consistent black hole growth and feedback \citep{dimatteoetal08}. They show that the black hole correlation function consists of two distinct components: contributions from intra-halo and inter-halo black hole pairs, i.e., the one-halo and two-halo terms. At small scales the one-halo term still follows a power law, which is different from that of dark matter. This boost in small scale power is due to galaxies hosting multiple black holes.
\begin{table}
\begin{center}
\begin{tabular}[t]{c|c|c|c|c}
\hline
\hline
\multicolumn{1}{c|}{Boxsize}&
\multicolumn{1}{c|}{$N_{p}$}&
\multicolumn{1}{c|}{$m_{\rm{DM}}$}&
\multicolumn{1}{c|}{$m_{\rm{gas}}$}&
\multicolumn{1}{c|}{$\epsilon$}\\
\multicolumn{1}{c|}{($h^{-1}$ Mpc)}&
\multicolumn{1}{c|}{}&
\multicolumn{1}{c|}{($h^{-1}M_{\odot}$)}&
\multicolumn{1}{c|}{($h^{-1}M_{\odot}$)}&
\multicolumn{1}{c|}{($h^{-1}$ kpc)}\\
\hline
 $33.75$ & $2\times486^{3}$ & $2.75\times10^{7}$ & $4.24\times10^{6}$ & $2.73$  \\ 
\hline
\hline 
\end{tabular}
\caption{The numerical parameters in the simulation. $ N_{p}$, $ m_{\rm{DM}}$, $m_{\rm{gas}}$ and $\epsilon$ are defined as the total number of particles, mass of the dark matter particles, mass of the gas particles, and comoving gravitational softening length respectively.}
\end{center}
\end{table}

Despite their advantages, studies using hydrodynamical simulations are limited due to their computational costs. For example simulations are limited to small boxes \citep{dimatteoetal08} restricting the analysis to small scales and low luminosity AGN. The computational cost of running these simulations to $z=0$ limits the possibility to compare the outputs with observations in the local universe. Also since every modification of the recipies adopted to describe the `sub-grid' physics requires rerun of the simulations, it is not feasible to do systematic parameter studies. Semi-analytic studies are advantageous over hydro simulations in this context. We now employ the SPH simulation of \citet {dimatteoetal08} to investigate the relation between AGN and their host halos within the halo occupation distribution (HOD) framework, a useful analytic formalism for modeling and interpreting AGN clustering. In our approach, the expensive hydrodynamic simulation is used to obtain insights on a general analytic technique for studying AGN clustering. The HOD \citep[e.g.,][] {m&f00, seljak00, b&w02} provides an analytic formalism for understanding clustering properties of galaxies or, in this case, AGN. It is characterized by the probability $P(N|M)$ that a halo of mass $M$ contains $N$ AGN of a given type, together with their spatial and velocity distribution inside halos. As long as the HOD at a fixed halo mass is statistically independent of the large scale environments of halos \citep[e.g.,][]{bondetal91}, it provides a complete description of the relation between AGN and halos, allowing the calculation of any clustering statistics (e.g., two point correlation function, three point correlation function, void probability distribution, pairwise velocity distribution) at all scales (small, intermediate, large) for a given cosmological model. This implies that if we can constrain the HOD empirically, we will have knowledge of everything that the measured clustering properties have to tell us about AGN formation and evolution models. The HOD allows to distinguish between the background cosmology and AGN evolution models. The cosmological parameters are encoded in the distribution of halos, where as the bias between mass and AGN is fully described in the probability distribution $P(N|M)$ (see \citealt{b&w02}, \citealt{z&w07} for the strength of HOD). The HOD formalism has been widely used in interpreting galaxy clustering \citep[see][]{zehavietal05}. 

In this paper we perform a theoretical study on the relation between AGN and dark matter halos using cosmological simulations. We assess the validity of several simplifying assumptions in the HOD modeling of AGN and its predictive power and limitations for modeling AGN clustering data. Our paper is organized as follows: In \S 2 we describe our simulations and give a brief description of our methodology. In \S 3 we present the conditional luminosity function of AGN, the mean occupation distribution, and the radial profiles of AGN. Finally in \S 4 we summarize our main conclusions and provide future directions to this work.
 
\begin{figure*}
\begin{center}
 \begin{tabular}{c}
      \includegraphics[width=16cm, height=6cm]{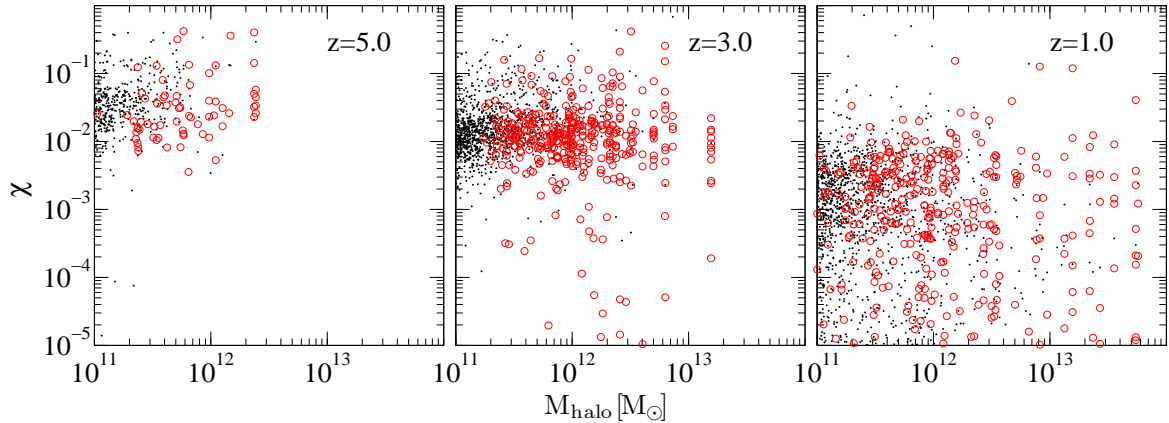} 
\end{tabular}
  \caption{Distribution of accretion rates scaled by the Eddington value ($\chi = \dot{M}/\dot{M}_{\rm {EDD}}$) as a function of halo mass at redshift $5.0$ (left panel), $3.0$ (middle panel), and $1.0$ (right panel). The black points show the distribution of the central AGN while the red open circles show the distribution of the satellite AGN. The mean rate depends weakly on halo mass at all redshifts. For a given halo mass, AGN at higher redshifts accrete at higher rates than at low redshifts. }

\end{center}
 \end{figure*} 

\section{Simulation}

\subsection{Numerical Code}
The numerical code uses a standard $\Lambda$CDM cosmological model with cosmological parameters from the first year WMAP results \citep{spergeletal03}. The simulation uses an extended version of the parallel cosmological Tree Particle Mesh-SPH code GADGET2 \citep{springel05}. Gas dynamics is modeled with Lagrangian SPH \citep{monaghan92}; radiative cooling and heating processes are computed from the prescription given by \citet{katzetal96}. The relevant physics of star formation and the associated supernova feedback have been approximated based on a sub-resolution multiphase model for the interstellar medium developed by
\citet{s&h03}. The size of the simulation box is $33.75h^{-1}$ comoving Mpc with periodic boundary conditions. A detailed description of the implementation of black hole accretion and the associated feedback model is given in \citet{dimatteoetal08}.  Black holes are
represented as collisionless ``sink" particles that can grow in mass by accreting gas or by mergers. The Bondi-Hoyle relation \citep{bondi52} is used to model the accretion rate of gas onto a black hole and capture the environmental dependence of black hole accretion. The accretion rate is given by $\dot{M}_{\rm{BH}} =
4\pi[G^{2}M_{\rm{BH}}^{2}\rho]/(c_{s}^{2} + v^{2})^{3/2}$, where $\rho$ and $c_{s}$ are density and speed of sound of the local gas, $v$ is the velocity of the black hole with respect to the gas, and $G$ is gravitational constant. Although the Bondi parameterization assumes spherical accretion, observations show that the Bondi relation adequately captures the physical state of the black hole and the Bondi scaling holds at scales much larger than the Bondi radius \citep[e.g.,][]{allenetal06}.

The bolometric luminosities of the AGN are $L_{\rm{Bol}} = \eta\dot{M}_{\rm{BH}}c^{2}$ where $\eta=0.1$ is the canonical efficiency for thin disk accretion. It is assumed that a small fraction of the bolometric luminosity couples to the surrounding gas as feedback energy $E_{f}$ such that \(\dot{E_{f}} = \epsilon_{f}L_{\rm{Bol}}\) with the feedback efficiency $\epsilon_f$ taken to be $5\%$. Based on previous studies of galaxy mergers, the value of the feedback efficiency parameter $\epsilon_f$ is chosen to be 5\% in the simulation, a necessary fraction to reproduce the normalization of the observed $M_{\rm{BH}}-\sigma$ relation at $z=0$ \citep{dimatteoetal05}. This feedback energy is put directly into the gas smoothing kernel at the position of the black hole \citep{dimatteoetal08}. The feedback energy is assumed to be distributed isotropically for the sake of simplicity; however the response of the gas can be anisotropic \citep[e.g.,][]{chatetal08, dimatteoetal08}.  The formation mechanism for the seed black holes which evolve into the observed supermassive black holes today is not known. The simulation creates seed black holes in halos that cross a specified mass threshold. At a given redshift, halos are defined by a friends-of-friends group finder algorithm
run on the fly. For any halo with mass $M \geq 5\times 10^{10}h^{-1}M_\odot$ that does not contain a black hole, the densest gas particle is converted to a black hole of mass $M_{\rm{BH}}=5\times 10^{5}h^{-1}M_\odot$. The choice of the seed mass in the present simulation was based on current models of seed black hole formation in the universe \citep[e.g.,][]{b&l03, b&l04}. The black hole then grows via the accretion prescription given above and by mergers with other black holes \citep{dimatteoetal08}. The simulation parameters are shown in Table 1.  
 
The detailed parameter studies of this simulation, and comparison with several observations have been previously done by \citet {sijackietal07, dimatteoetal08, c&d08, croftetal09, degrafetal10}. The model has reproduced the observed $ M_{\rm{BH}}-\sigma$ relation, total black hole mass density \citep{dimatteoetal08}, and the AGN luminosity functions and their evolution in optical, soft and hard X-ray band \citep{degrafetal10}. The black hole mass density from the simulation matches well with the constraints from the integrated X-ray luminosity function \citep{shankaretal04, marconietal04} and the accretion rate density is consistent with the constraints of \citet{hopkinsetal07b}. This model has been also used in galaxy merger simulations to investigate black hole growth and their correlation with host galaxies \citep{dimatteoetal05, robertsonetal06, hopkinsetal07a}.  

The model has some intrinsic limitations. It does not treat the physics of the accretion disc in detail, which is not feasible in a simulation with cosmological volume. Nevertheless, the model is capable of representing some key aspects of black hole evolution in a cosmological context and in reproducing many observational results. Although the model has some limitations, it has been fairly successful in reproducing some of the key observational results. Several other teams \citep{b&s09, johanssonetal08, teyssieretal11} have now implemented similar modeling for black hole accretion and feedback in hydrodynamic simulations. \citet{b&s09} have independently explored the parameter space of the fiducial model of \citet{dimatteoetal08}. \citet{teyssieretal11} have implemented this prescription in an adaptive mesh refinement code (different from the SPH formalism) and showed that AGN feedback can solve the overcooling problem in clusters. This large number of previous investigations make this particular subgrid model a good choice for studying co-evolution of AGN with their host dark matter halos. We emphasize that in this work we do not intend to explore the parameter space related to the subgrid modeling of black holes. Instead we study the relation between AGN and dark matter halos in one simulation box. The goal of our current study is to obtain a useful description between AGN and dark matter to guide the interpretation of observations (e.g., AGN clustering) and to learn about AGN evolution using the model of \citet{dimatteoetal08}.

\subsection{Sample Selection}
For our analysis we consider halos above a mass scale of $10^{11}M_{\odot}$ and black holes with masses above $10^{6}M_{\odot}$. This choice of mass scales will minimize the effect of the seed mass in our results. We also restricted our analysis to black holes with Eddington scaled accretion rate $\chi = \dot{M}/\dot{M}_{\rm {EDD}} \geq 10^{-5}$ since we want to study systems with active accretion. This selection imposes a luminosity cut of $10^{38}$ ergs/s in our sample.
One of the key goals of this paper is to separate the contributions to the occupation distribution from the central and satellite AGN (as done for galaxy HOD by \citet {zhengetal05}). We consider all the black holes within $R_{200}$ (defined as the radius within which the enclosed mean density is $200$ times the critical density) of a halo to be associated with the host halo. The most massive black hole within $R_{200}$ is assumed to be the central AGN and the rest are designated as satellites. At $z = 1.0, 3.0,$ and $5.0$ about $79\%$, $93\%$, and $91\%$ of the central black holes lie within $20\%$ of $R_{200}$ respectively. We expect the central black hole to be located at the center of the main galaxy in the halo since this is the place of highest gas density and thus maximal black hole growth. The fact that most `central AGN' are found near the group center shows that our technique for identifying the central AGN is reasonably adequate. Due to the limited volume of our simulation we could only probe the faint end of the AGN luminosity function \citep{degrafetal10}. For calculating the mean occupation we study AGN samples with bolometric luminosities above $10^{42}$ ergs/s, corresponding to the observational limit of X-ray selected AGN in deep surveys \citep[e.g.,][]{uedaetal03, boyleetal93, cowieetal03}. We also show the distribution of AGN with luminosities above $10^{40}$ ergs/s and $10^{38}$ ergs/s for theoretical interests to see the luminosity dependence of HOD parameters.
\begin{figure*}
\begin{center}
 \begin{tabular}{c}
       \includegraphics[width=16cm]{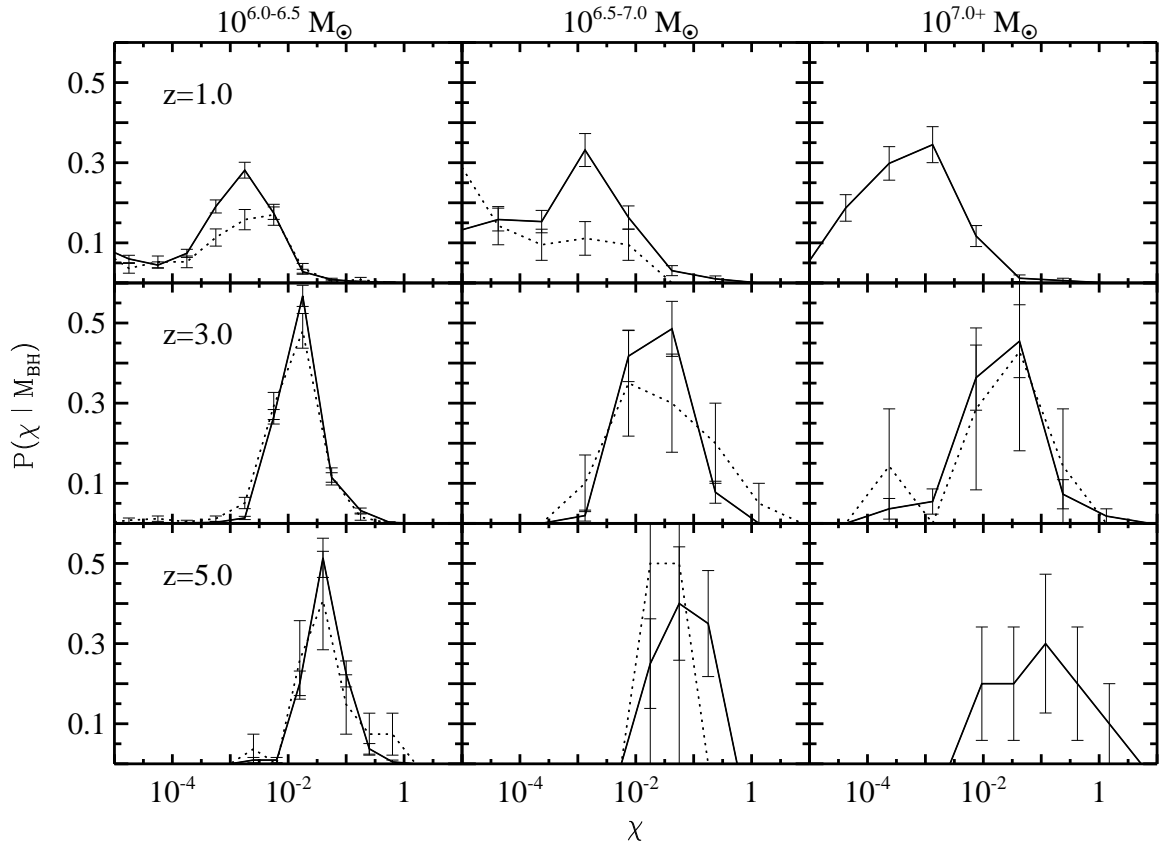} 
 \end{tabular}
  \caption{The distribution of Eddington scaled accretion rates ($\chi$) for different black hole mass and redshift bins. From left to right, columns correspond to black hole mass bins, $6.0 \leq \rm {Log}(M_{\rm{BH}}) \leq 6.5$, $6.5 \leq \rm {Log}(M_{\rm{BH}}) \leq 7.0$, and $ \rm{Log}(M_{\rm{BH}})\geq 7.0$. The top, middle, and bottom panels correspond to redshifts $1.0$, $3.0$, and $5.0$ respectively. The solid and dotted lines represent the central and satellite distributions. The mean of the distribution is lower at $z=1.0$ compared to redshifts $3.0$ and $5.0$ for all mass bins. The error bars are Poisson error bars.}
\end{center}
 \end{figure*} 


In a companion paper, \citet{degrafetal11b} (hereafter Paper I), we use the same SPH simulation to characterize the HOD of black holes as a function of black hole mass. Our present study is fundamentally different from that in Paper I. We aim to study the HOD of AGN in terms of AGN luminosity. By definition the HOD is always studied based on physical properties of the object. In the case of galaxy clustering, the galaxy HOD has been extensively studied in the literature based on color, luminosity, stellar mass (see \citealt{zehavietal05}, \citealt{zhengetal07}). Black hole mass and luminosity (accretion rates) are different parameters and there does not exist an obvious one-to-one correspondence between mass and accretion rates (e.g. Fig.\ 1). Additional baryonic physics is needed to go from black hole mass to accretion rate (and thus luminosity). There is some correlation between mass and accretion rates, but the scatter is huge and the correlation depends strongly on redshift \citep{c&d08,chatetal08}. 
\begin{figure*}
\begin{center}
 \begin{tabular}{c}
       \includegraphics[angle=0.0, width=16cm]{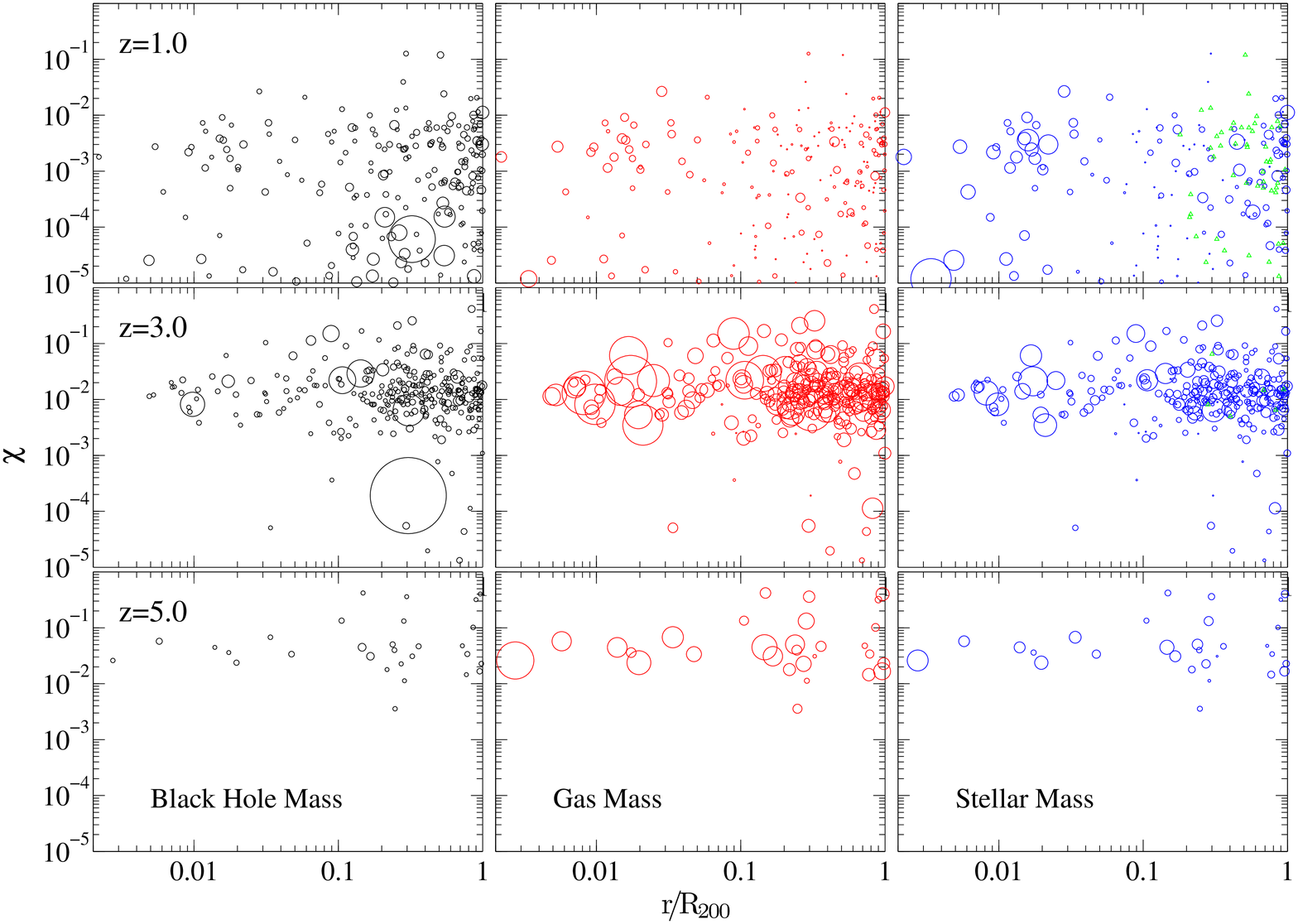}
\end{tabular}
  \caption{Spatial distribution of the satellite AGN. The size of each circle is proportional to the mass of the black hole (left panel), gas mass (middle panel), and stellar mass (right panel). The gas and the stellar mass are computed within a spherical region of $10$~kpc around the black hole. The green triangles represent black holes with no stars within $10$ kpc. The top, middle and bottom panels correspond to $z=1.0$, $z=3.0$, and $z=5.0$ respectively. The distribution has been shown for all satellite AGN residing in halos with $M_{\rm{halo}} \geq 10^{11}M_{\odot}$ and black hole mass $M_{\rm{BH}} \geq 10^{6}M_{\odot}$. The accretion rates do not show any dependence on radius. The accretion rates do not show any pattern with black hole mass either, but there is a correlation with gas mass and stellar mass. In general the accretion rate is higher for a higher gas (stellar) mass. As we go to low redshifts the gas mass and stellar mass around black holes decreases due to the effect of AGN feedback. AGN feedback is responsible for expelling gas from the vicinity of the black hole and suppressing its own growth.}
\end{center}
 \end{figure*} 

In practice, it is barely possible to have accurate mass measurement of a large sample of black holes beyond the local universe. Therefore, the study in Paper I is mostly of theoretical interest. By investigating the occupation properties of black holes at the subhalo level, Paper I develops an analytic mechanism to populate dark matter halos with black holes in simulation. On the contrary, AGN luminosity is more readily measured observationally and there are existing and ongoing efforts in measuring AGN/quasar clustering as a function of luminosity. Our study in this paper will provide the necessary ingredient in interpreting these observations. Similar to the mass function of black holes and luminosity function of AGN, the black-hole-mass based HOD and AGN-luminosity based HOD are complementary tools to advance our understanding of black hole evolution.

Since the goal of this paper is to provide the HOD framework to interpret observations we adopt some technical differences in the definition of dark matter halos and the choice of our sample from Paper I. In Paper I we differentiate between central and satellite black holes at the subhalo level. Black holes residing in the central subgroup are called central black holes and black holes residing in satellite subgroups are called satellite black holes. In the nomenclature in Paper I there can be multiple central black holes. We emphasize that the model in Paper I will be useful for populating halos (N body simulations) with black holes using semi-analytic approaches and for studying the distribution of black holes in cosmological simulations. On the other hand our approach in this paper is to perform a theoretical study on the relation between AGN (selected based on luminosity) and dark matter halos and examine the assumptions in the HOD for modeling AGN clustering data. Our identification of central (i.e. most massive black hole within $R_{200}$) and satellite (rest of the black holes within $R_{200}$) AGN is equivalent to primary and secondary black holes in Paper I (see Fig.\ 2 of Paper I) and is similar to the standard terminology used for galaxy HOD. We note that the friends-of-friends halos were used in Paper I for studying black hole occupation. Here we choose $R_{200}$ to define the halo boundary, which is commonly adopted in defining galaxy clusters. Since the HOD describes bias at the level of systems near dynamical equilibrium, traditional virial methods of mass estimation can take advantage of this and provide a direct measurement of the probability distribution $P(N|M)$ \citep{b&w02}. We have verified that the two different halo definitions show a good agreement in assigning black holes to host halos within $10\%$.



\begin{figure*}
\begin{center}
 \begin{tabular}{c}
       \includegraphics[width=16cm]{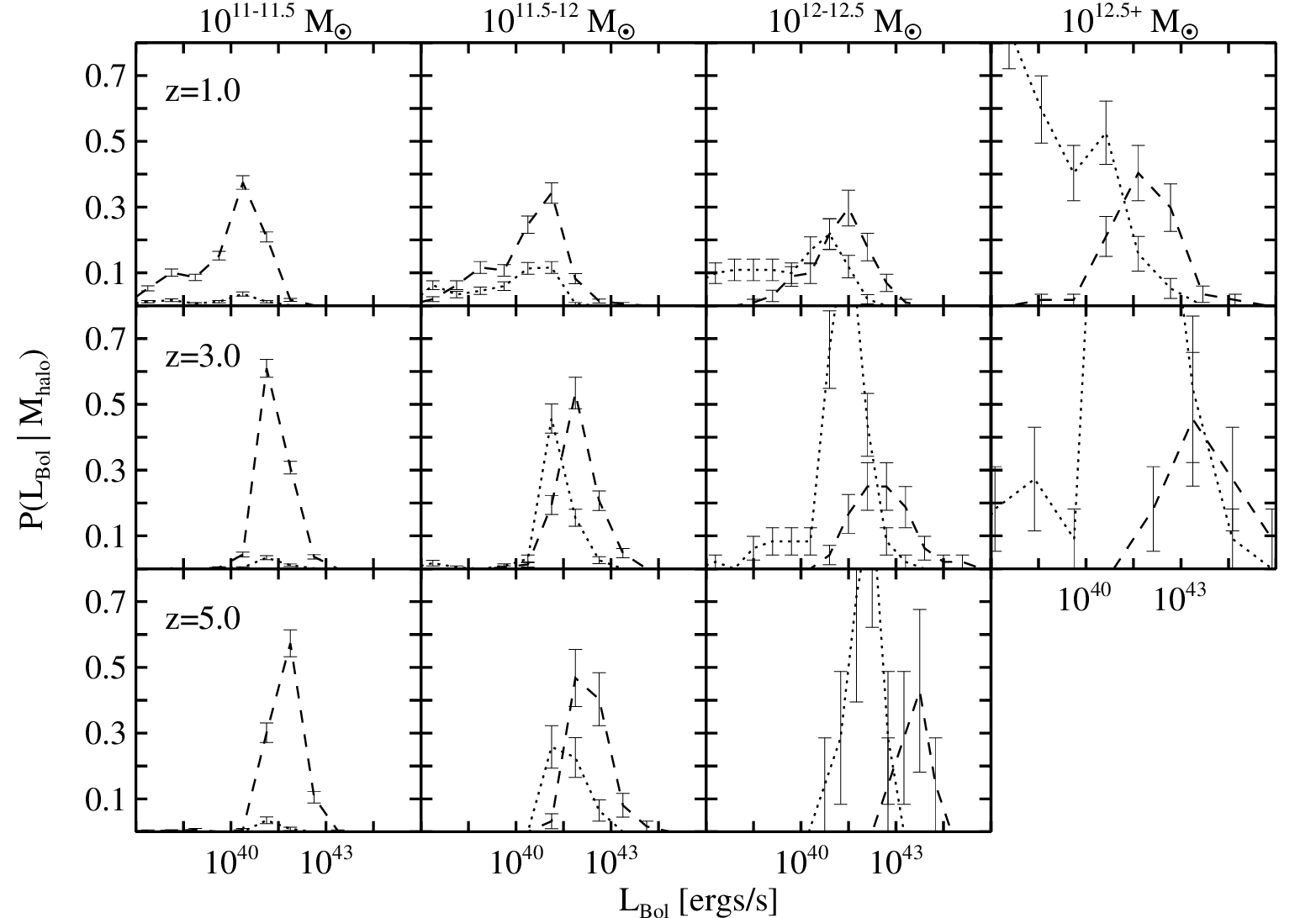}
\end{tabular}
  \caption{The conditional luminosity functions of the AGN in the simulation. The top, middle and bottom panels show the probability distributions for redshifts $1.0$, $3.0$, and $5.0$, respectively. The halo mass bins range from $10^{11}M_{\odot}-10^{11.5}M_{\odot}$ (leftmost), to $\geq 10^{12.5}M_{\odot}$ (rightmost). The dashed and dotted lines represent the probability distribution for the central and satellite AGN. The curves are normalized by the total number of halos in each mass bin. The central AGN traces a log-normal distribution as commonly assumed in semi-analytic studies. The error bars represent the Poisson error bars in each bin.}
\end{center}
 \end{figure*}

\section{Results}
In this section we present the distribution of accretion rates as a function of halo mass and black hole mass. From this we calculate the conditional luminosity functions and the mean occupation distribution of central and satellite AGN. Finally we show the radial distribution of the satellite AGN and provide the best-fit hod parameters. 

\subsection{The Accretion Rates}
Figure 1 shows the distribution of the Eddington scaled accretion rate ($\chi = \dot{M}/\dot{M}_{\rm {EDD}}$), as a function of halo mass for three redshifts. The black points are central AGN and the red open circles represent satellites. The mean rate is roughly independent of halo mass, which is commonly assumed in semi-analytic studies. However, the mean value of $\chi$ at a given halo mass varies with redshift. AGN at higher redshifts (left panel of Fig.\ 1 showing $z=5.0$) tend to accrete at a higher value than their low redshift counterparts (right panel of Fig.\ 1 showing $z=1.0$). Figure 2 shows the distribution of $\chi$ as a function of black hole mass. It is defined as the conditional probability distribution of $\chi$ for a given black hole mass $P(\chi|M_{\rm{BH}})$. The solid and the dotted lines denote central and satellite AGN, respectively. The mean rates are again roughly independent of black hole mass but vary substantially with redshift. The central and satellite distributions tend to trace each other at all redshifts.

In Fig.\ 3 we present the spatial distribution of the satellite AGN. The sizes of the circles are proportional to the black hole mass (left panel), gas mass (middle panel), and stellar mass (right panel) respectively. The gas and the stellar mass are computed within a spherical region of radius $10$~kpc surrounding the black hole. The green triangles represent black holes with zero stellar mass within $10$ kpc. The top, middle, and bottom panels are for $z=1.0$, $3.0$, and $5.0$ respectively. The AGN accretion rates do not show any dependence on radius or black hole mass. Also we do not see any correlation between black hole mass and its location within the halo. The trends of accretion rates with gas and stellar mass are more prominent. The low values of $\chi$ at low redshift do not correlate with the mass distribution of the black holes (left panel of Fig.\ 3). The mass distribution of black holes is not significantly different at different redshifts. At low redshifts the gas mass around black holes decreases and so do the accretion rates. This is due to the effect of AGN feedback. Outflow from AGN is responsible for expelling gas from the vicinity of the black hole and suppressing its own growth. AGN feedback is responsible for shutting down star-formation too. In Fig.\ 3 we also see the evidence of suppressed stellar mass in the immediate neighborhood of the AGN. We see evidence of how a local phenomenon (feedback) affects the global distribution of AGN, regardless of the host halo mass (see Fig.\ 1 for the host halo masses). This picture is consistent with what has been seen in previous studies \citep[e.g.,][]{dimatteoetal08, sijackietal07} and semi-analytic models \citep[e.g.,][]{w&l03}

\subsection{The Conditional Luminosity Function}
The conditional luminosity function (CLF) \citep[e.g.,][] {yangetal03}, in this case for AGN, is defined as the luminosity distribution $\Phi(L|M)$ of AGN that reside in halos of a given mass $M$. The global luminosity function is given by 
\begin{equation}
\Phi(L) = \int \frac{dn}{dM}\Phi(L|M)dM,
\end{equation}
where $\Phi(L)$ is the AGN luminosity function and $dn/dM$ is the halo mass function. The CLF constitutes the differential form of the HOD. The CLF provides a tool to examine the distribution of halo mass for a given luminosity and study luminosity dependent clustering. This formalism has been widely used in modeling galaxy clustering with data from 2dFGRS and DEEP2 redshift surveys \citep[e.g.,][]{vandenboschetal03}. The CLF is useful in generating mock catalogs which can then be used to test the cosmological model \citep[e.g.,][] {moetal04, yanetal04}. Figure 4 shows the CLF in bins of halo mass running from Log $M = 11.25 \pm 0.25$ (left most panel) to Log $M \geq 12.5$ (right most panel). The dashed lines show the distribution of bolometric luminosities for the central AGN while the dotted lines show the distributions for the satellite AGN. In both cases the distribution has been normalized by the total number of halos in the given mass bin. The top, middle, and bottom panels show results for redshifts $1.0$, $3.0$, and $5.0$, respectively. 

The luminosity distributions reveal several features. The central AGN luminosity distribution closely traces a log-normal distribution \citep[e.g.,][]{mw01}. Fig.\ 5 captures the features we observe in the CLF. The top panel of Fig.\ 5 shows the evolution of the mean log luminosity of the central AGN as a function of halo mass. The bottom panel of Fig.\ 5 shows the $1 \sigma$ scatter in the log of the luminosity. The mean luminosity at a given halo mass evolves with redshift, with higher mean luminosity at higher redshifts. The scatter in the luminosity distribution of central AGN increases toward lower redshift, indicating that the luminosities for a given halo mass are more uniform at high redshift than at low redshift. As a result of this, at low redshifts AGN samples based on luminosity will have a wider range of host halo masses and hence clustering will depend weakly on luminosity. However at high redshifts luminosity dependent clustering will be more prominent. Similar redshift evolution of luminosity dependent clustering has been observed with SDSS quasars \citep{shenetal09, shenetal07}. We note that the clustering properties of AGN will be largely related to their host halo mass. However we find that accretion rates and hence luminosity depends strongly on local properties particularly at low redshifts (e.g., feedback discussed in \S 3.1) which erases some of the dependences of bolometric luminosity on halo mass and hence the luminosity dependent clustering at low redshifts. This will not be the case in black hole mass selected sample since black hole mass is more tightly correlated with halo mass regardless of redshifts \citep{c&d08}. The shape of the satellite luminosity function cannot be identified definitively due to lack of statistics. We note that the lower end of the satellite luminosity function is affected by the resolution of the simulation and the seed black hole mass, which is manifested in the cut-off of the luminosity function. The satellite distribution also shows some variation with halo mass. For a given halo mass the peak of the satellite distribution tends to be at a lower luminosity than the central AGN. 

\begin{figure}
\begin{center}
 \begin{tabular}{c}
        \includegraphics[angle=0.0,height=10cm, width=8cm]{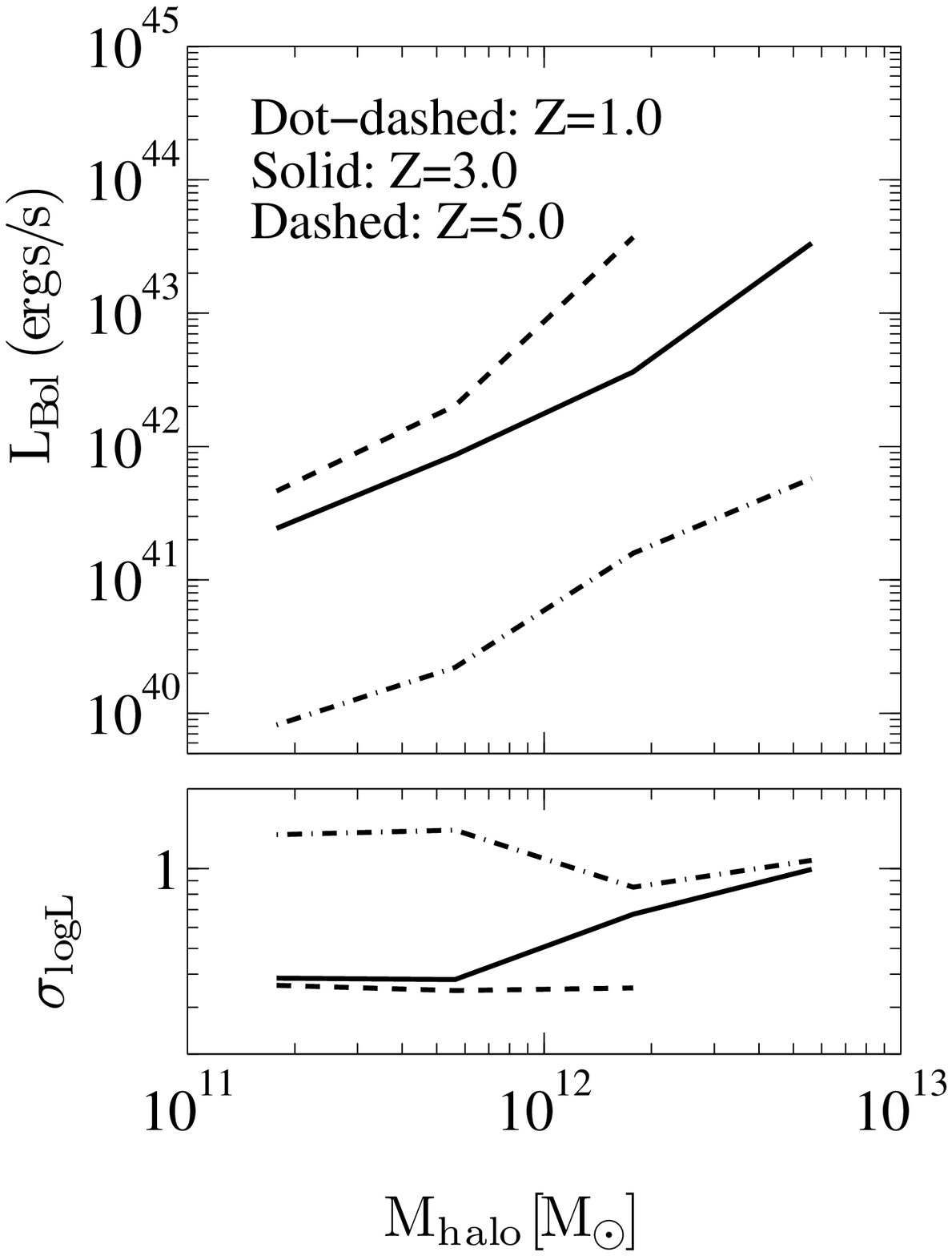}
\end{tabular}
\caption{The mean (upper panel) and scatter (lower panel) of the central AGN conditional luminosity function as a function of halo mass, shown in Figure 4. The scatter is defined as the $1\sigma$ scatter from the mean in Log$L_{\rm{Bol}}$. The dot-dashed, solid, and dashed lines denote redshifts $1.0$, $3.0$, and $5.0$, respectively. For a given halo mass the mean increases with redshifts. The growth of the black holes is possibly suppressed and hence the accretion rate (and the luminosity thereof) is lower at lower redshifts. The scatter in the distribution is also high at low redshifts since several factors (e.g., feedback from black holes) introduce spread in the $L_{\rm{Bol}}-M_{\rm{halo}}$ relation.}
\end{center}
 \end{figure} 

\begin{figure}
\begin{center}
 \begin{tabular}{c}
       \includegraphics[width=8cm]{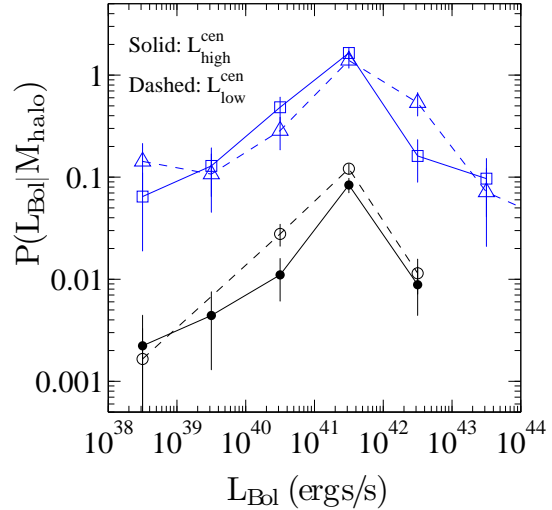}
\end{tabular}
  \caption{Dependence of the conditional luminosity function of satellite AGN on central AGN luminosity. The open and the filled black circles connected by the dashed and solid lines respectively represent the halo mass bin $11.0\leq {\rm Log}(M_{{\rm halo}}) \leq 12.0$. Similarly the blue open triangles and the open squares connected by the dashed and solid lines respectively represent the halo mass bin  $12.0\leq {\rm Log}(M_{{\rm halo}}) \leq 13.0$. The solid and the dashed lines represent the distribution of satellite AGN luminosities for central AGN with $L_{{\rm Bol}}$ above (higher) and below (lower) the mean luminosity (see Fig.\ 5) respectively. We do not see any significant correlation between central AGN luminosity and the number of satellites. The error bars at each bin represent the Poisson error bars.}
\end{center}
 \end{figure}
\begin{figure}
\begin{center}
 \begin{tabular}{c}
       \resizebox{80mm}{!}{\includegraphics[angle=0.0]{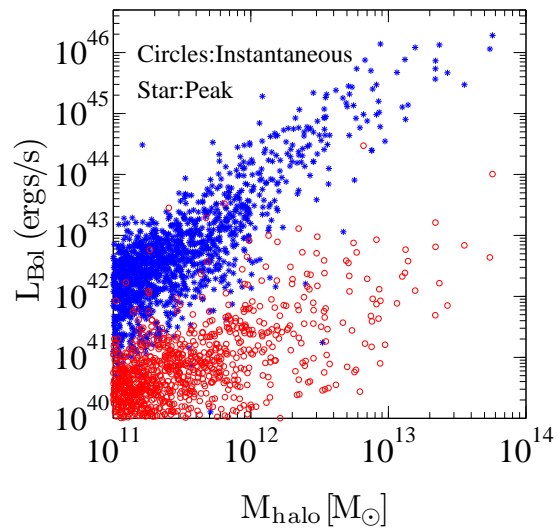}} 
\end{tabular}
  \caption{Bolometric luminosity of the central AGN as a function of halo mass calculated at $z=1.0$. The blue points show the peak luminosity between $z=2.0$ and $z=1.0$, and the red open circles show the instantaneous luminosity at $z=1.0$. The peak luminosity tends to correlate more tightly with halo mass than instantaneous luminosity.}
\end{center}
 \end{figure}   

To examine the effect of the central AGN on the number distribution of satellites within a halo, we compare the conditional luminosity functions of satellite AGN for halos differing in central AGN luminosities. This is defined as the conditional distribution of satellite AGN luminosities for a fixed $M_{\rm halo}$ and $L_{{\rm Bol}}^{{\rm cen}}$. We used a large halo mass bin to increase the statistics of our sample. We note that there is a correlation between central AGN luminosity and halo mass. To eliminate the effect of mass-dependent central AGN luminosity, we divide halos according to the central AGN luminosity as follows. For each redshift, at each halo mass, we tag halos as `high Lcen' (`low Lcen') if their central AGN luminosities are above (below) the mean central AGN luminosity at that mass (Fig.5). Then the conditional luminosity functions of satellite AGN are computed for the `high Lcen' and `low Lcen' halos, respectively. The results are shown in Fig.\ 6. The solid lines show the distribution of satellite luminosities for `high Lcen' sample and the dashed lines represent the distribution of satellite luminosities for `low Lcen' sample. 

The two groups of curves are for two halo mass bins. The open and the filled black circles connected by the dashed and solid lines respectively represent the halo mass bin $11.0\leq {\rm Log}(M_{{\rm halo}}) \leq 12.0$. Similarly the blue open triangles and the open squares connected by the dashed and solid lines respectively represent the halo mass bin  $12.0\leq {\rm Log}(M_{{\rm halo}}) \leq 13.0$. The curves are normalized to the total number of halos hosting the central AGN above and below the mean luminosity. We have shown the result for $z=3.0$. Our results are similar for $z=1.0$ and $z=5.0$. The correlation between central and satellite AGN luminosity at a fixed halo mass will have important implications on the small scale clustering strength. If there exists a strong large scale feedback effect from the central AGN it can possibly alter the local gas distribution around satellite AGN and suppress the growth of the satellite black holes. This in effect will decrease the number of luminous satellites in a halo with a higher central AGN luminosity and we would observe an anti-correlation between satellite number and central AGN luminosity. 

In Fig.\ 6 we do not see any correlation between the number of satellites and the luminosity of the central AGN. It has been shown in \citet{chatetal08} that in group scale halos (the most massive halos in our simulation) feedback from the central AGN can extend up to a few hundred kpc and thus can potentially affect the satellite distribution. However in higher mass halos satellite AGN will also be residing in more massive subhalos and so they will be relatively unaffected from the feedback effects of the central AGN. We note that in the self regulatory growth paradigm the AGN shuts down its own growth by blowing up the gas around it (as seen in Fig.\ 3) but the feedback effects do not affect the gas distribution around neighboring black holes.

\subsection{Peak versus Instantaneous Luminosity}
\begin{figure*}
\begin{center}
 \begin{tabular}{c}
       \includegraphics[width=16cm]{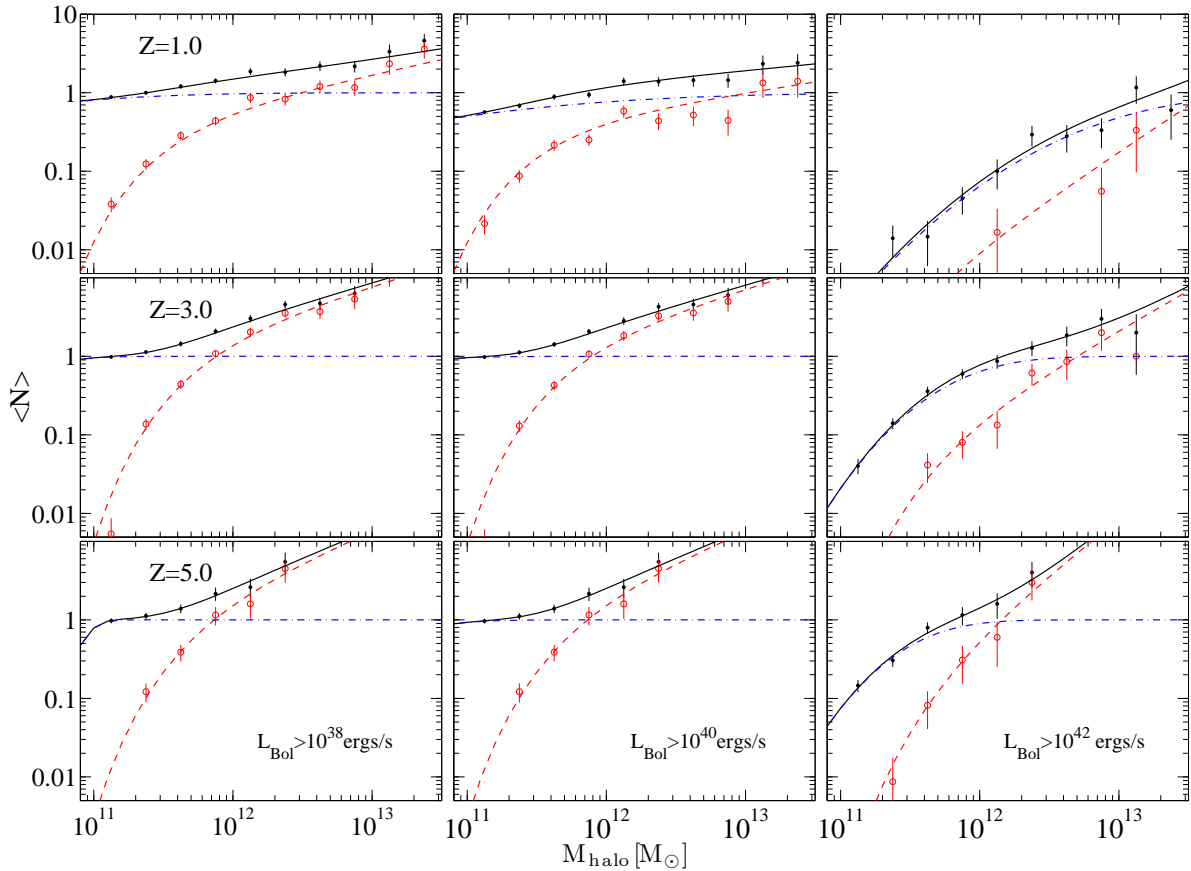} 
 \end{tabular}
  \caption{The mean occupation distribution of the AGN as a function of halo mass. The top, middle, and bottom panels correspond to redshifts $1.0$, $3.0$, and $5.0$, respectively. The left, middle and right columns correspond to different luminosity cuts. The black points represent the mean occupation of all the AGN within $R_{200}$ and the red open circles show the contribution from the satellite AGN. The mean occupancy of the central AGN can be idealized as a softened step function while the satellite population can be approximated by a power law (see Eqs.\ 4 and 5). The black solid, blue dot-dashed, and the red dashed lines are the best-fit models for the total, central, and satellite occupation respectively. The best-fit HOD parameters are shown in Table 2. The error bars reflect the $1 \sigma$ Poisson error bars.}
\end{center}
 \end{figure*}   
We see a difference in the mean of the CLF for the central AGN between $z=1.0$ and $z=3.0$. To investigate this effect we extracted the light curves of the AGN \citep{c&d08, degrafetal10} and looked at their peak luminosities between $z=2.0$ and $z=1.0$. The result is shown in Fig.\ 7. The peak luminosities (blue stars) of the central AGN show a lower scatter with halo mass and the best-fit relation is 
\begin{equation}
L_{\rm{Bol}}=10^{45}\left(\frac{M_{\rm{halo}}}{10^{12.93}M_{\odot}}\right)^{1.65 \pm 0.07} {\rm ergs/s},
\end{equation} 
where $L_{\rm{Bol}}$ is the peak bolometric luminosity. We overplot the instantaneous luminosity at $z=1.0$ (red open circles). The best-fit slope for the instantaneous luminosity is $(0.95 \pm 0.12)$. As we go to higher redshifts the AGN have luminosities closer to their peak luminosities and hence we see a higher mean value and a tighter correlation (Fig.\ 5). The difference between the peak and the instantaneous luminosity is most prominent at higher mass halos. We associate this suppression with feedback from AGN. In Paper I, we used the $M_{\rm{BH}} -\sigma$ relation to show that black holes residing in higher mass halos enter the feedback dominated phase at an earlier time than black holes populating lower mass halos. Thus feedback effects will alter the $M_{\rm{halo}}-L_{\rm{Bol}}$ correlation resulting in a wider distribution of host halo masses for a given AGN luminosity at lower redshift. This result is also in agreement with \citet{lidzetal06} who conclude that the peak luminosity is more correlated with halo mass than the instantaneous luminosity of AGN and hence a better indicator of clustering. However, we cannot measure the peak luminosity of the AGN in practice. 




\subsection{The Mean Occupation Function of AGN}
 We model the mean occupation function of AGN in dark matter halos by decomposing it into a more physically illuminating central and satellite contributions \citep[e.g.,][] {kravtsovetal04, zhengetal05, zehavietal05}
\begin{equation}
\langle N\left(M\right)\rangle = \langle N_{\rm{cen}}\left(M\right)\rangle +\langle N_{\rm{sat}}\left(M\right)\rangle
\end{equation}
where $\langle N_{\rm{cen}} \left(M\right)\rangle$ represents the mean occupation function of central AGN and $\langle N_{\rm{sat}}\left(M\right)\rangle$ represents the mean occupation function of satellites (see discussions in \S 2.2 for the differences in terminology with Paper I). This formalism implicitly assumes that the AGN content of halos of a given mass is statistically independent of the large-scale environments within which the halos reside and thus the mean occupation $\langle N \left(M\right)\rangle$ depends only on halo mass  \citep[e.g.,][]{bondetal91,l&k99}. $\langle N \left(M\right)\rangle$ is shown as a function of halo mass in Fig.\ 8. The black filled circles show the total occupation and the red open circles show the satellite occupation. The black solid, blue dot-dashed, and the red dashed lines are the best-fit models for the total, central, and satellite occupation. The top, middle, and bottom panels show redshifts $1.0$, $3.0$, and $5.0$, respectively, while the left, middle, and right columns denote luminosity thresholds $10^{38}$ ergs/s, $10^{40}$ ergs/s, and $10^{42}$ ergs/s. 

We see that the central AGN occupation number follows a distribution close to a softened step function and the satellite occupation follows a power law similar to the galaxy or the dark matter subhalo case \citep[e.g.,][] {kravtsovetal04, zhengetal05}. Our HOD model is defined as follows:
\begin{eqnarray}
\langle N_{\rm{cen}}\rangle &=& \frac{1}{2}\left[1+{\rm erf}\left(\frac{{\rm Log} M-{\rm Log} M_{\rm{min}}}{\sigma_{\rm{Log M}}}\right)\right]\\
 \langle N_{\rm{sat}}\rangle &=& \mbox{} (M/M_{1})^{\alpha}\exp(-M_{\rm{cut}}/M) 
\end{eqnarray} 
In this formalism there are four parameters for modeling the HOD: $M_{\rm{min}}$, defining the halo mass where the occupation of central AGN of a given type (in the present case we choose AGN in terms of luminosity type) is $0.5$; $\sigma_{\rm{Log M}}$ the characteristic transition width; $M_{1}$, the mass scale at which the mean number of satellites above a given luminosity threshold equals unity; and $\alpha$, the power law exponent for the satellite occupation. However it has been observed that the mean occupation of the satellites drops off faster than a power law at lower halo mass \citep [e.g.,][]{zhengetal05, kravtsovetal04, conroyetal06} and we also see this trend in our case. We use the parameterization of \citet{tinkeretal05} to model this drop-off. The parameter $M_{\rm{cut}}$ is used to model the rolling off the power law. So we have a five parameter HOD model. The best-fit HOD parameters are shown in Table 2. The change in the luminosity threshold does affect the mean value of the power law exponent but the values are consistent within $1\sigma$. The power law exponent also show weak evolution (seen in Paper I) with redshift but is consistent with no evolution within statistical limits. 
\begin{figure}
\begin{center}
 \begin{tabular}{c}
       
         \includegraphics[width=8cm]{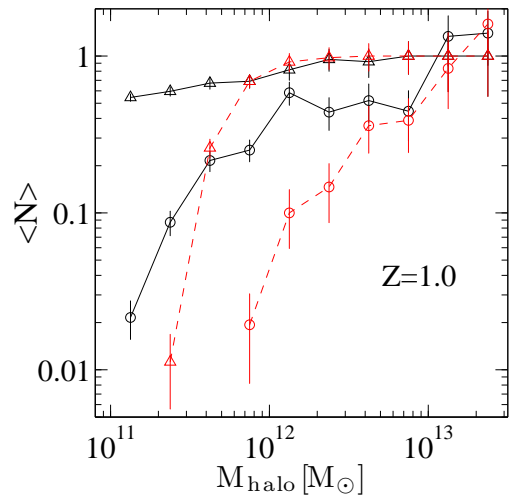}
 \end{tabular}
  \caption{The mean occupation distribution of the AGN for different selections. The black solid lines represent the mean occupation function of AGN with bolometric luminosities greater than $10^{40}$ ergs/s at $z=1.0$ (shown in the top middle panel of Fig.\ 8). The triangles represent the mean occupation of central AGN and the open circles represent the mean occupation of satellite AGN. We now use the best-fit relation between black hole mass and AGN luminosity at $z=1.0$ \citep{chatetal08} to obtain the equivalent mass corresponding to our bolometric luminosity of $10^{40}$ ergs/s. The red dashed lines represent the mean occupation functions for the mass selected sample with triangles representing central black holes and open circles representing satellite black holes. In all cases the error bars are $1 \sigma$ Poisson error bars.}
\end{center}
 \end{figure}   

We compute the distribution of AGN with respect to the mean $ P(N|\langle N \rangle)$. In Paper I we showed that the distribution of satellite number (without any mass or luminosity cut) follows a Poisson distribution. We have verified that this is also the case for AGN in any luminosity threshold sample. We further performed a Kolmogorov-Smirnov (KS) test to check whether the satellite distribution follows a Poisson distribution in our luminosity selected sample. The mean P-value that we obtain by performing the KS test over all redshifts and over all mass and luminosity bins is $0.92$. This shows that the null hypothesis is strongly accepted and the distribution of the satellites is close to a Poisson distribution.

 \begin{table*}
\begin{center}
\begin{tabular}[t]{c|c|c|c|c|c|c|c|c}
\hline
\hline
\multicolumn{1}{c|}{Redshift}&
\multicolumn{1}{c|}{$L_{\rm{Bol}}$(ergs s$^{-1}$)}&
\multicolumn{1}{c|}{$N_{{\rm tot}}^{{\rm cen}}$}&
\multicolumn{1}{c|}{$N_{{\rm tot}}^{{\rm sat}}$}&
\multicolumn{1}{c|}{Log$(M_{\rm{min}}/M_{\odot})$}&
\multicolumn{1}{c|}{$\sigma_{\rm{LogM}}$}&
\multicolumn{1}{c|}{$\alpha$}&
\multicolumn{1}{c}{Log$(M_{1}/M_{\odot})$}&
\multicolumn{1}{c}{Log$(M_{\rm{cut}}/M_{\odot})$}\\
\hline
       &$\geq 10^{38}$ &$1248$ &$336$ &$10.12 \pm 0.15$ &$1.45 \pm 0.21$  &$0.38 \pm 0.05$ & $12.38\pm 0.63$  & $11.50$\\ 
 $1.0$ &$\geq 10^{40}$ &$927$  &$226$ &$11.00 \pm 0.11$ &$1.98 \pm 0.20$  &$0.27 \pm 0.08$ & $12.98\pm 1.32$  & $11.50$\\ 
       &$ \geq 10^{42}$ &$56$   &$4$   &$13.03 \pm 0.27$ &$0.96 \pm 0.27$  &$1.17 \pm 1.12$ & $13.64\pm 9.01$ & $11.50$\\
\hline 
       &$\geq 10^{38}$ &$1123$ &$380$ &$10.41 \pm 0.06$ &$0.50 \pm 0.05$ &$0.55 \pm 0.13$ & $11.36\pm 1.48$ & $11.70$\\ 
 $3.0$ &$\geq 10^{40}$ &$1119$ &$361$ &$10.05 \pm 0.08$ &$0.78 \pm 0.13$ &$0.54 \pm 0.13$ & $11.39\pm 1.41$ & $11.70$\\ 
       &$\geq 10^{42}$ &$199$  &$42$  &$11.85 \pm 0.05$ &$0.59 \pm 0.06$ &$1.00 \pm 0.32$ & $12.66\pm 2.12$ & $11.70$\\ 
\hline
       &$\geq 10^{38}$ &$410$  &$66$  &$10.91 \pm 0.04$ &$0.16 \pm 0.03$ &$0.75 \pm 0.18$ & $11.47\pm 2.81$ & $11.70$\\ 
 $5.0$ &$\geq 10^{40}$ &$407$  &$66$  &$10.41 \pm 0.06$ &$0.57 \pm 0.05$ &$0.75 \pm 0.18$ & $11.47\pm 2.81$ & $11.70$\\ 
       &$\geq 10^{42}$ &$121$  &$18$   &$11.53 \pm 0.09$ &$0.52 \pm 0.18$ &$1.37 \pm 0.37$ & $12.05\pm 3.83$ & $11.70$\\ 
\hline 
\hline
\end{tabular}
\end{center}
\caption{AGN HOD parameters for three redshifts corresponding to Eqs.\ 4 and 5. Columns 3 and 4 show the total number of central and satellite black holes in each bin.}
\end{table*}
\subsection{Luminosity versus Mass}
Our model of central occupation is slightly different than Paper I. In Paper I we do not impose any mass cut and hence the fraction of halos (above the threshold mass) containing black holes is always unity (see Eqs.\ 1 and 2 in Paper I). In this paper the softening of the step function arises from the luminosity based selection. We note that at lower luminosities the duty cycle is extremely close to $1.0$ at all halo massbins and we converge to the model of Paper I. However the drop-off from unity is evident in the $10^{42}$ ergs/s sample (right panel of Fig.\ 8). In Fig.\ 9 we show the differences in the mean occupation function between a mass selected and a luminosity selected sample. The black solid lines show the mean occupation function of AGN with bolometric luminosities greater than $10^{40}$ ergs/s at $z=1.0$ (shown in top middle panel of Fig.\ 8). The triangles represent the mean occupation of central AGN and the open circles represent the mean occupation of satellite AGN. We now use the best-fit relation between black hole mass and AGN luminosity at $z=1.0$ \citep{chatetal08} to obtain the equivalent mass corresponding to our bolometric luminosity of $10^{40}$ ergs/s. The red dashed lines represent the mean occupation functions for the mass selected sample with triangles representing central black holes and open circles representing satellite black holes. 

We see a clear difference between the two populations at lower halo masses. For the central AGN (triangles) the solid and the dashed lines converge at a mass scale of $10^{11.5}M_{\odot}$. Also the mass-selected occupation function falls-off very steeply below this mass scale. This difference arises from lower mass black holes residing in lower mass halos with high Eddington ratios. Also the correspondence between black hole mass and host halo mass is tighter than luminosity and hence we see this sharp cut-off in the occupation function for the mass selected sample. Recently \citet{galloetal10} found evidence of such low mass high accretion rate black holes in local early type galaxies in the AMUSE-Virgo survey. For the satellite population (open circles) the lack of convergence in the occupation function between the two population is prominent even at higher halo masses. The minimum halo mass for hosting satellite black holes (based on mass) is much higher than minimum halo mass for hosting satellite AGN with equivalent luminosity. We thus see that the HOD properties will be significantly different between a mass selected sample and a luminosity selected sample. This is because a relatively small halo can host a bright AGN if there is high density gas. On the other hand, it is difficult to get a massive black hole in a small halo, since that would require significant amounts of dense gas over a long period of time.


\subsection{Radial Profiles}
Fig.\ 9 shows the radial distribution of the number density of satellite AGN (luminosity greater than $10^{40}$ ergs/s) within host halos of masses $11.0\leq {\rm Log}(M_{{\rm halo}}) \leq 13.0$. The red circles, blue triangles, and the green squares show the profile at $z=1.0$, $z=3.0$, and $z=5.0$ respectively. It has been shown in Paper I that the radial distribution of black holes follow a power law at all redshifts (Eq.\ 6 of Paper I). Our findings are similar to Paper I. The black dashed line is the best-fit power law to the average profile over all redshifts (since we do not see any significant evolution with redshift). The best-fit values for $\beta$ (power law index of AGN: averaged over all redshifts) and ${\rm Log}(n_{0})$ (normalization: averaged over all redshifts) are $-2.33 \pm 0.08$ and $-0.67 \pm 0.05$ for the sample shown in Fig.\ 9. The power law index $\beta$ does not show any significant dependence on halo mass or AGN luminosity either. For comparison we also show the profile of dark matter and galaxies within AGN host halos. The solid and the dotted lines show the average profiles (averaged over all redshifts) for dark matter and satellite galaxies respectively. The minimum stellar mass that we used for obtaining the galaxy profile is $10^{8}M_{\odot}$ (roughly $100$ times the stellar mass resolution). The profiles do not show any variation if we change the threshold stellar mass for selecting galaxies. The current choice of stellar mass is an optimization between resolution elements and statistics. The profiles are normalized to the mean number density of the corresponding species (satellite AGN/dark matter/ satellite galaxies) within $R_{200}$ \citep[e.g.,][]{n&k05}.


We also fit NFW profile \citep{nfw97} to the AGN radial distribution. We fit the profiles for different concentration parameters and calculate the corresponding P values. At all redshifts our data strongly disfavors the null hypothesis and the NFW profile is ruled out at $3\sigma$. We note that the AGN are more centrally concentrated than dark matter and galaxies. The reason that we see an enhanced population of AGN at the center of the halo compared to galaxies is because of the merging process of black holes. When two galaxies merge there exists a time lag between merging galaxies and the merging of AGN that reside in them. This time lag is due to the time it takes for the satellite AGN to fall in to the halo center where it can merge with the central AGN. After the two galaxies merge the time that it takes for the AGN to merge can then be further affected by the gas content of these galaxies. \citet{l&m07} measure the radial profile of radio sources in clusters and show that it is consistent with an NFW profile with a concentration of $25$. \citet{martinietal07} study the radial distribution of X-ray selected AGN in clusters and find that AGN with X-ray luminosities above $10^{42}$ ergs/s show stronger central concentration than cluster host galaxies. However, a bigger sample with ${\rm L_{X}} \geq 10^{41}$  ergs/s (closer to our sample of AGN) does not show any stronger evidence of central concentration; different from what we observe in simulation. 

In Fig.\ 11 we show the relation between stellar mass and AGN luminosity at $z=1.0$. The stellar mass is computed within a spherical region of radius $30$~kpc surrounding the black hole. We selected different spatial scales to compute the stellar mass. The values converge between radii of $25$ kpc to $30$ kpc and hence we chose $30$ kpc to be the relevant radius for computing the stellar mass content in AGN host galaxies. The black points represent central AGN and the red open circles show the satellite distribution. In the case of central AGN the stellar mass for more luminous AGN is generally higher. AGN will have higher luminosity when the accretion rate is high, which is more likely when the gas density is high. Also when the gas density is high the cooling rate will be higher and hence there will be more stars. So in general higher density will correspond to high accretion rates of AGN and higher stellar mass. For satellite AGN this trend is very weak. But even for satellite AGN if we consider a luminosity selected sample we will systematically choose AGN with higher stellar mass. Also in Fig.\ 3 we see that AGN with higher stellar mass content tend to reside in the central regions of the halos. This can potentially increase the higher concentration of AGN seen in Fig.\ 10 for a luminosity selected sample. To further check this we selected AGN based on their stellar mass content and computed their average radial profile. We find a steeper mean slope ($\beta \sim -2.7$) for the stellar mass selected sample compared to the luminosity selected sample. This again is an indication of enhanced central concentration of AGN with higher stellar mass content in their host galaxies.


\section{Summary and Discussions}
In this paper, we study the relation between AGN and dark matter halos using a cosmological SPH simulation that incorporates black hole and galaxy formation physics. Specifically we have investigated the effect of environmental factors (e.g., feedback, local gas density, host galaxy mass) on the accretion rates of AGN within dark matter halos and how it affects the occupation distribution of AGN. We examine a number of simplifying assumptions in the HOD modeling (e.g., dependence of AGN luminosity on halo mass, dependence of AGN accretion on black hole mass) of AGN and provide the necessary tools to model AGN clustering data. We have characterized the HOD of faint AGN (the sample probed by our simulation box) which can be generalized to incorporate the bright end of the luminosity function. 

\begin{figure}
\begin{center}
 \begin{tabular}{c}
       \includegraphics[width=8cm]{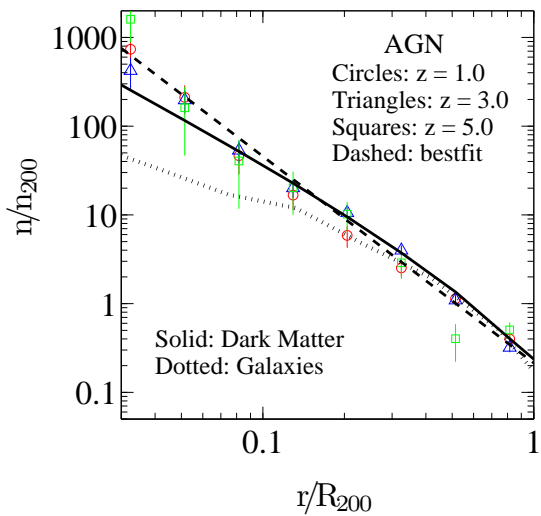}\\
      
 \end{tabular}
\caption{The radial profile of satellite AGN. The open circles, triangles, and squares show the profiles at $z=1.0$, $z=3.0$, and $z=5.0$ respectively. The profiles are obtained by averaging over AGN with $L_{\rm{Bo}l}>10^{40}$ ergs/s and host halo masses between $10^{11}-10^{13}M_{\odot}$. The radial profiles do not exhibit any significant evolution with redshift. The dashed line shows the average power-law fit over all redshifts. The best-fit value for the slope is $-2.33 \pm 0.08$. For comparison we also show the profile of dark matter and galaxies within AGN host halos. The solid and the dotted lines show the average profiles (averaged over all redshifts) for dark matter and satellite galaxies respectively. The error bars are representative of the $1 \sigma$ error bars.}
\end{center}
 \end{figure} 

We compute the conditional luminosity functions of AGN and separate the contributions from the central and satellite AGN. Our key findings are as follows. (1) The central AGN luminosities follow a log-normal distribution similar to the assumption in semi-analytic studies. (2) The mean of the CLF for a given halo mass shows a strong dependence on redshift with higher luminosities at higher redshifts. (3) The scatter in central AGN luminosity is large at low redshift, but decreases with increasing redshift. This implies that the dependence of AGN clustering on luminosity is weak at low redshift, but it can become stronger at high redshift. We analyze the light curves of individual AGN and show that there exists a tighter correlation between halo mass and peak luminosity rather than instantaneous luminosity. We present the joint distribution of satellite occupation as a function of halo mass and central AGN luminosity. We do not see any significant correlation between the satellite number and the luminosity of the central AGN. We also show that the mean occupation function of the central AGN resembles a softened step function while the satellite population follows a power law with an exponential roll-off at lower mass, similar to what has been observed with the galaxy HOD. We show that low mass black holes with high Eddington ratios residing in low mass halos makes the luminosity based HOD significantly different from the black hole mass based HOD.

We will now compare our simulation results with semi-analytic models that have been widely used for clustering analysis \citep[e.g.,][]{mw01, shankaretal10}. We compare our results with the model described in \citet{shankaretal10} who computed the mass function of black holes at $z=3.0$ using the halo mass function and a redshift dependent relation between halo mass and black hole mass. They parameterize the redshift dependence in the normalization of the black hole mass-halo mass relation while the slope is kept fixed at $1.52$. In our simulation we observe the average slope to be $(1.04\pm 0.06)$ at $z=3.0$. There is a large scatter from the mean slope and the $1\sigma$ scatter is $0.44$. We note that this relation in our simulation is dependent on the AGN model and the ratio of the threshold halo mass that can host black holes to the mass of the seed black holes. \citet{shankaretal10} assumes the black holes to be accreting with a constant Eddington fraction. Although the mean $\chi$ in our simulation is roughly independent of halo mass as seen in Figure 1, they do show an evolution with redshift. This has been also noted in \citet{shankaretal10} where they suggest that the assumption of constant Eddington fraction might break down at the faint end. The other important parameter is the scatter in the $L_{\rm{Bol}}-M_{\rm{halo}}$ relation. We find a log-normal distribution as assumed in semi-analytic studies. However the scatter strongly depends on redshift with increasing scatter at low redshifts (shown in Figure 5). 


Observationally the most relevant physical quantity describing an AGN is luminosity. Recently \citet{miyajietal10} and \citet{starikovaetal10} have used X-ray selected AGN to constrain the HOD empirically. We propose to compare our HOD model with observational samples in a follow-up paper. However an alternative approach would be selecting black holes based on their mass and measure clustering statistics based on black hole mass. Although we do not have reliable observational measurements of black hole masses, we have also provided the framework for predicting clustering properties based on the black hole mass function in Paper I. These two complimentary approaches would impose even tighter constraints on theoretical models of AGN growth and feedback. Our analysis provides the first step toward comparing semi-analytic and simulation results on AGN clustering and assessing some of the simplifying assumptions in the present interpretations of AGN clustering observations and constraining physical parameters. Because of the small simulation box, our current study is limited to low luminosity AGN, and the results on satellite AGN suffer from small number statistics. A larger simulation box with larger statistical samples spanning both the faint and the bright end of the luminosity function can provide tighter constraints on AGN HOD. Also, the mass of the seed black hole is dependent on the resolution of the simulation, imposing an artificial mass cut in the simulation. Future work on high resolution simulation and more accurate modeling of accretion and feedback \citep[e.g.,][]{b&s09} will also be needed to understand the full implication of gas physics and black hole accretion on clustering studies. The HOD formalism has been successfully incorporated in galaxy evolution. We hope that our work will have the same impact for studying the co-evolution of AGN and their hosts and shed light on the AGN-galaxy connection.  

\begin{figure}
\begin{center}
 \begin{tabular}{c}
       \includegraphics[width=8cm]{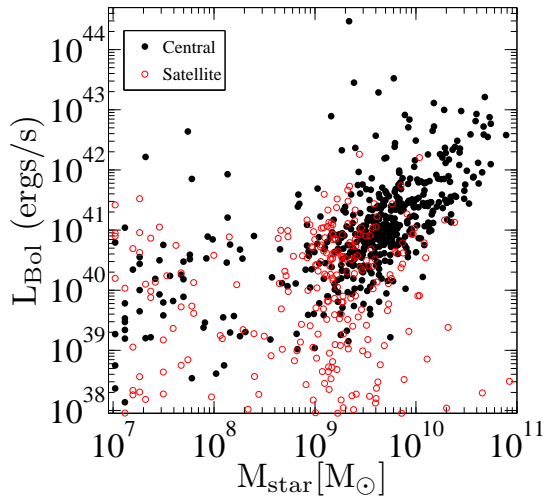}\\
      
 \end{tabular}
\caption{Relation between stellar mass and AGN luminosity at $z=1.0$. The stellar mass is computed within a spherical region of radius $30$~kpc surrounding the black hole. The black points represent central AGN and the red open circles show the satellite distribution. In the case of central AGN the stellar mass for more luminous AGN is generally higher. For satellite AGN this trend is very weak. But even for satellite AGN if we consider a luminosity selected sample we will systematically choose AGN with higher stellar mass. }
\end{center}
 \end{figure} 

\section*{Acknowledgments}

We would like to thank Douglas Rudd, Laurie Shaw, and Frank van den Bosch for some useful discussions which helped in the analysis and physical interpretation of some results. We thank David Weinberg for suggesting the analysis for looking at the correlation between central and satellite AGN luminosities. We also thank the referee for the comments which helped in improving the draft. JR was supported by the Yale College Dean's Science Research Fellowship through summer 2010. ZZ gratefully acknowledges support from Yale Center for Astronomy and Astrophysics through a YCAA fellowship. DN was supported in part by the NSF grant AST-1009811, by NASA ATP grant NNX11AE07G, and by Yale University.  The work in CMU was supported by the National Science Foundation, NSF Petapps, OCI-0749212 and NSF AST-1009781.
The simulations were carried out at the NSF Teragrid Pittsburgh Supercomputing Center (PSC).
 
\bibliography{mybib}{}

\begin{thebibliography}{84}
\expandafter\ifx\csname natexlab\endcsname\relax\def\natexlab#1{#1}\fi

\bibitem[{{Allen} {et~al.}(2006){Allen}, {Dunn}, {Fabian}, {Taylor}, \&
  {Reynolds}}]{allenetal06}
{Allen} S.~W., {Dunn} R.~J.~H., {Fabian} A.~C., {Taylor} G.~B., {Reynolds}
  C.~S., 2006, \mnras, 372, 21

\bibitem[{{Berlind} \& {Weinberg}(2002)}]{b&w02}
{Berlind} A.~A., {Weinberg} D.~H., 2002, \apj, 575, 587

\bibitem[{{Bond} {et~al.}(1991){Bond}, {Cole}, {Efstathiou}, \&
  {Kaiser}}]{bondetal91}
{Bond} J.~R., {Cole} S., {Efstathiou} G., {Kaiser} N., 1991, \apj, 379, 440

\bibitem[{{Bondi}(1952)}]{bondi52}
{Bondi} H., 1952, \mnras, 112, 195

\bibitem[{{Bonoli} {et~al.}(2009){Bonoli}, {Marulli}, {Springel}, {White},
  {Branchini}, \& {Moscardini}}]{bonolietal09}
{Bonoli} S., {Marulli} F., {Springel} V., {White} S.~D.~M., {Branchini} E.,
  {Moscardini} L., 2009, \mnras, 396, 423

\bibitem[{{Booth} \& {Schaye}(2009)}]{b&s09}
{Booth} C.~M., {Schaye} J., 2009, \mnras, 398, 53

\bibitem[{{Boyle} {et~al.}(1993){Boyle}, {Griffiths}, {Shanks}, {Stewart}, \&
  {Georgantopoulos}}]{boyleetal93}
{Boyle} B.~J., {Griffiths} R.~E., {Shanks} T., {Stewart} G.~C.,
  {Georgantopoulos} I., 1993, \mnras, 260, 49

\bibitem[{{Boyle} {et~al.}(2000){Boyle}, {Shanks}, {Croom}, {Smith}, {Miller},
  {Loaring}, \& {Heymans}}]{boyleetal00}
{Boyle} B.~J., {Shanks} T., {Croom} S.~M., {Smith} R.~J., {Miller} L.,
  {Loaring} N., {Heymans} C., 2000, \mnras, 317, 1014

\bibitem[{{Bromm} \& {Larson}(2004)}]{b&l04}
{Bromm} V., {Larson} R.~B., 2004, \araa, 42, 79

\bibitem[{{Bromm} \& {Loeb}(2003)}]{b&l03}
{Bromm} V., {Loeb} A., 2003, \apj, 596, 34

\bibitem[{{Cattaneo} {et~al.}(2006){Cattaneo}, {Dekel}, {Devriendt},
  {Guiderdoni}, \& {Blaizot}}]{cattaneoetal06}
{Cattaneo} A., {Dekel} A., {Devriendt} J., {Guiderdoni} B., {Blaizot} J., 2006,
  \mnras, 370, 1651

\bibitem[{{Chatterjee} {et~al.}(2008){Chatterjee}, {Di Matteo}, {Kosowsky}, \&
  {Pelupessy}}]{chatetal08}
{Chatterjee} S., {Di Matteo} T., {Kosowsky} A., {Pelupessy} I., 2008, \mnras,
  390, 535

\bibitem[{{Colberg} \& {Di Matteo}(2008)}]{c&d08}
{Colberg} J.~M., {Di Matteo} T., 2008, \mnras, 387, 1163

\bibitem[{{Conroy} {et~al.}(2006){Conroy}, {Wechsler}, \&
  {Kravtsov}}]{conroyetal06}
{Conroy} C., {Wechsler} R.~H., {Kravtsov} A.~V., 2006, \apj, 647, 201

\bibitem[{{Cowie} {et~al.}(2003){Cowie}, {Barger}, {Bautz}, {Brandt}, \&
  {Garmire}}]{cowieetal03}
{Cowie} L.~L., {Barger} A.~J., {Bautz} M.~W., {Brandt} W.~N., {Garmire} G.~P.,
  2003, \apjl, 584, L57

\bibitem[{{Croft} {et~al.}(2009){Croft}, {Di Matteo}, {Springel}, \&
  {Hernquist}}]{croftetal09}
{Croft} R.~A.~C., {Di Matteo} T., {Springel} V., {Hernquist} L., 2009, \mnras,
  400, 43

\bibitem[{{Croom} {et~al.}(2004){Croom}, {Boyle}, {Shanks}, {Outram}, {Smith},
  {Miller}, {Loaring}, {Kenyon}, \& {Couch}}]{croometal04}
{Croom} S., {Boyle} B., {Shanks} T., {Outram} P., {Smith} R., {Miller} L.,
  {Loaring} N., {Kenyon} S., {Couch} W., 2004, in Astronomical Society of the
  Pacific Conference Series, Vol. 311, AGN Physics with the Sloan Digital Sky
  Survey, {G.~T.~Richards \& P.~B.~Hall}, ed., pp. 457--+

\bibitem[{{Croton} {et~al.}(2006){Croton}, {Springel}, {White}, {De Lucia},
  {Frenk}, {Gao}, {Jenkins}, {Kauffmann}, {Navarro}, \&
  {Yoshida}}]{crotonetal06}
{Croton} D.~J., {Springel} V., {White} S.~D.~M., {De Lucia} G., {Frenk} C.~S.,
  {Gao} L., {Jenkins} A., {Kauffmann} G., {Navarro} J.~F., {Yoshida} N., 2006,
  \mnras, 365, 11

\bibitem[{{Degraf} {et~al.}(2010){Degraf}, {Di Matteo}, \&
  {Springel}}]{degrafetal10}
{Degraf} C., {Di Matteo} T., {Springel} V., 2010, \mnras, 402, 1927

\bibitem[{{Degraf} {et~al.}(2011{\natexlab{a}}){Degraf}, {Di Matteo}, \&
  {Springel}}]{degrafetal11a}
{Degraf} C., {Di Matteo} T., {Springel} V., 2011{\natexlab{a}}, \mnras, 413, 1383

\bibitem[{{Degraf} {et~al.}(2011{\natexlab{b}}){Degraf}, {Oborski}, {Di
  Matteo}, {Chatterjee}, {Nagai}, {Richardson}, \& {Zheng}}]{degrafetal11b}
{Degraf} C., {Oborski} M., {Di Matteo} T., {Chatterjee} S., {Nagai} D.,
  {Richardson} J., {Zheng} Z., 2011{\natexlab{b}}, \mnras, 416, 1591

\bibitem[{{Di Matteo} {et~al.}(2008){Di Matteo}, {Colberg}, {Springel},
  {Hernquist}, \& {Sijacki}}]{dimatteoetal08}
{Di Matteo} T., {Colberg} J., {Springel} V., {Hernquist} L., {Sijacki} D.,
  2008, \apj, 676, 33

\bibitem[{{Di Matteo} {et~al.}(2005){Di Matteo}, {Springel}, \&
  {Hernquist}}]{dimatteoetal05}
{Di Matteo} T., {Springel} V., {Hernquist} L., 2005, \nat, 433, 604

\bibitem[{{Fan} {et~al.}(2001){Fan}, {Narayanan}, {Lupton}, {Strauss}, {Knapp},
  {Becker}, {White}, {Pentericci}, {Leggett}, {Haiman}, {Gunn}, {Ivezi{\'c}},
  {Schneider}, {Anderson}, {Brinkmann}, {Bahcall}, {Connolly}, {Csabai}, {Doi},
  {Fukugita}, {Geballe}, {Grebel}, {Harbeck}, {Hennessy}, {Lamb}, {Miknaitis},
  {Munn}, {Nichol}, {Okamura}, {Pier}, {Prada}, {Richards}, {Szalay}, \&
  {York}}]{fanetal01}
{Fan} X., {Narayanan} V.~K., {Lupton} R.~H., {Strauss} M.~A., {Knapp} G.~R.,
  {Becker} R.~H., {White} R.~L., {Pentericci} L., {Leggett} S.~K., {Haiman} Z.,
  {Gunn} J.~E., {Ivezi{\'c}} {\v Z}., {Schneider} D.~P., {Anderson} S.~F.,
  {Brinkmann} J., {Bahcall} N.~A., {Connolly} A.~J., {Csabai} I., {Doi} M.,
  {Fukugita} M., {Geballe} T., {Grebel} E.~K., {Harbeck} D., {Hennessy} G.,
  {Lamb} D.~Q., {Miknaitis} G., {Munn} J.~A., {Nichol} R., {Okamura} S., {Pier}
  J.~R., {Prada} F., {Richards} G.~T., {Szalay} A., {York} D.~G., 2001, \aj,
  122, 2833

\bibitem[{{Gallo} {et~al.}(2010){Gallo}, {Treu}, {Marshall}, {Woo}, {Leipski},
  \& {Antonucci}}]{galloetal10}
{Gallo} E., {Treu} T., {Marshall} P.~J., {Woo} J.-H., {Leipski} C., {Antonucci}
  R., 2010, \apj, 714, 25

\bibitem[{{Gebhardt} {et~al.}(2000){Gebhardt}, {Bender}, {Bower}, {Dressler},
  {Faber}, {Filippenko}, {Green}, {Grillmair}, {Ho}, {Kormendy}, {Lauer},
  {Magorrian}, {Pinkney}, {Richstone}, \& {Tremaine}}]{gebhardtetal00}
{Gebhardt} K., {Bender} R., {Bower} G., {Dressler} A., {Faber} S.~M.,
  {Filippenko} A.~V., {Green} R., {Grillmair} C., {Ho} L.~C., {Kormendy} J.,
  {Lauer} T.~R., {Magorrian} J., {Pinkney} J., {Richstone} D., {Tremaine} S.,
  2000, \apjl, 539, L13

\bibitem[{{Gilli} {et~al.}(2005){Gilli}, {Daddi}, {Zamorani}, {Tozzi},
  {Borgani}, {Bergeron}, {Giacconi}, {Hasinger}, {Mainieri}, {Norman},
  {Rosati}, {Szokoly}, \& {Zheng}}]{gillietal05}
{Gilli} R., {Daddi} E., {Zamorani} G., {Tozzi} P., {Borgani} S., {Bergeron} J.,
  {Giacconi} R., {Hasinger} G., {Mainieri} V., {Norman} C., {Rosati} P.,
  {Szokoly} G., {Zheng} W., 2005, \aap, 430, 811

\bibitem[{{Graham} \& {Driver}(2007)}]{g&d07}
{Graham} A.~W., {Driver} S.~P., 2007, \mnras, 380, L15

\bibitem[{{Graham} {et~al.}(2011){Graham}, {Onken}, {Athanassoula}, \&
  {Combes}}]{grahametal11}
{Graham} A.~W., {Onken} C.~A., {Athanassoula} E., {Combes} F., 2011, \mnras,
  412, 2211

\bibitem[{{Hopkins} {et~al.}(2007{\natexlab{a}}){Hopkins}, {Hernquist}, {Cox},
  {Robertson}, \& {Krause}}]{hopkinsetal07b}
{Hopkins} P.~F., {Hernquist} L., {Cox} T.~J., {Robertson} B., {Krause} E.,
  2007{\natexlab{a}}, \apj, 669, 67

\bibitem[{{Hopkins} {et~al.}(2007{\natexlab{b}}){Hopkins}, {Richards}, \&
  {Hernquist}}]{hopkinsetal07a}
{Hopkins} P.~F., {Richards} G.~T., {Hernquist} L., 2007{\natexlab{b}}, \apj,
  654, 731

\bibitem[{{Hopkins} {et~al.}(2006){Hopkins}, {Robertson}, {Krause},
  {Hernquist}, \& {Cox}}]{hopkinsetal06}
{Hopkins} P.~F., {Robertson} B., {Krause} E., {Hernquist} L., {Cox} T.~J.,
  2006, \apj, 652, 107

\bibitem[{{Johansson} {et~al.}(2008){Johansson}, {Naab}, \&
  {Burkert}}]{johanssonetal08}
{Johansson} P.~H., {Naab} T., {Burkert} A., 2008, Astronomische Nachrichten,
  329, 956

\bibitem[{{Katz} {et~al.}(1996){Katz}, {Weinberg}, \& {Hernquist}}]{katzetal96}
{Katz} N., {Weinberg} D.~H., {Hernquist} L., 1996, \apjs, 105, 19

\bibitem[{{Kauffmann} \& {Haehnelt}(2000)}]{k&h00}
{Kauffmann} G., {Haehnelt} M., 2000, \mnras, 311, 576

\bibitem[{{Kravtsov} {et~al.}(2004){Kravtsov}, {Berlind}, {Wechsler}, {Klypin},
  {Gottl{\"o}ber}, {Allgood}, \& {Primack}}]{kravtsovetal04}
{Kravtsov} A.~V., {Berlind} A.~A., {Wechsler} R.~H., {Klypin} A.~A.,
  {Gottl{\"o}ber} S., {Allgood} B., {Primack} J.~R., 2004, \apj, 609, 35

\bibitem[{{Lapi} {et~al.}(2006){Lapi}, {Shankar}, {Mao}, {Granato}, {Silva},
  {De Zotti}, \& {Danese}}]{lapietal06}
{Lapi} A., {Shankar} F., {Mao} J., {Granato} G.~L., {Silva} L., {De Zotti} G.,
  {Danese} L., 2006, \apj, 650, 42

\bibitem[{{Lemson} \& {Kauffmann}(1999)}]{l&k99}
{Lemson} G., {Kauffmann} G., 1999, \mnras, 302, 111

\bibitem[{{Lidz} {et~al.}(2006){Lidz}, {Hopkins}, {Cox}, {Hernquist}, \&
  {Robertson}}]{lidzetal06}
{Lidz} A., {Hopkins} P.~F., {Cox} T.~J., {Hernquist} L., {Robertson} B., 2006,
  \apj, 641, 41

\bibitem[{{Lin} \& {Mohr}(2007)}]{l&m07}
{Lin} Y., {Mohr} J.~J., 2007, \apjs, 170, 71

\bibitem[{{Ma} \& {Fry}(2000)}]{m&f00}
{Ma} C., {Fry} J.~N., 2000, \apj, 543, 503

\bibitem[{{Malbon} {et~al.}(2007){Malbon}, {Baugh}, {Frenk}, \&
  {Lacey}}]{malbonetal07}
{Malbon} R.~K., {Baugh} C.~M., {Frenk} C.~S., {Lacey} C.~G., 2007, \mnras, 382,
  1394

\bibitem[{{Marconi} {et~al.}(2004){Marconi}, {Risaliti}, {Gilli}, {Hunt},
  {Maiolino}, \& {Salvati}}]{marconietal04}
{Marconi} A., {Risaliti} G., {Gilli} R., {Hunt} L.~K., {Maiolino} R., {Salvati}
  M., 2004, \mnras, 351, 169

\bibitem[{{Martini} {et~al.}(2007){Martini}, {Mulchaey}, \&
  {Kelson}}]{martinietal07}
{Martini} P., {Mulchaey} J.~S., {Kelson} D.~D., 2007, \apj, 664, 761

\bibitem[{{Martini} \& {Weinberg}(2001)}]{mw01}
{Martini} P., {Weinberg} D.~H., 2001, \apj, 547, 12

\bibitem[{{Marulli} {et~al.}(2009){Marulli}, {Bonoli}, {Branchini}, {Gilli},
  {Moscardini}, \& {Springel}}]{marullietal09}
{Marulli} F., {Bonoli} S., {Branchini} E., {Gilli} R., {Moscardini} L.,
  {Springel} V., 2009, \mnras, 396, 1404

\bibitem[{{Marulli} {et~al.}(2008){Marulli}, {Bonoli}, {Branchini},
  {Moscardini}, \& {Springel}}]{marullietal08}
{Marulli} F., {Bonoli} S., {Branchini} E., {Moscardini} L., {Springel} V.,
  2008, \mnras, 385, 1846

\bibitem[{{Menci} {et~al.}(2008){Menci}, {Fiore}, {Puccetti}, \&
  {Cavaliere}}]{mencietal08}
{Menci} N., {Fiore} F., {Puccetti} S., {Cavaliere} A., 2008, \apj, 686, 219

\bibitem[{{Merritt} \& {Ferrarese}(2001)}]{m&f01}
{Merritt} D., {Ferrarese} L., 2001, \apj, 547, 140

\bibitem[{{Miyaji} {et~al.}(2010){Miyaji}, {Krumpe}, {Coil}, \&
  {Aceves}}]{miyajietal10}
{Miyaji} T., {Krumpe} M., {Coil} A.~L., {Aceves} H., 2010, ArXiv e-prints

\bibitem[{{Mo} {et~al.}(2004){Mo}, {Yang}, {van den Bosch}, \&
  {Jing}}]{moetal04}
{Mo} H.~J., {Yang} X., {van den Bosch} F.~C., {Jing} Y.~P., 2004, \mnras, 349,
  205

\bibitem[{{Monaghan}(1992)}]{monaghan92}
{Monaghan} J.~J., 1992, \araa, 30, 543

\bibitem[{{Myers} {et~al.}(2006){Myers}, {Brunner}, {Richards}, {Nichol},
  {Schneider}, {Vanden Berk}, {Scranton}, {Gray}, \& {Brinkmann}}]{myersetal06}
{Myers} A.~D., {Brunner} R.~J., {Richards} G.~T., {Nichol} R.~C., {Schneider}
  D.~P., {Vanden Berk} D.~E., {Scranton} R., {Gray} A.~G., {Brinkmann} J.,
  2006, \apj, 638, 622

\bibitem[{{Nagai} \& {Kravtsov}(2005)}]{n&k05}
{Nagai} D., {Kravtsov} A.~V., 2005, \apj, 618, 557

\bibitem[{{Navarro} {et~al.}(1997){Navarro}, {Frenk}, \& {White}}]{nfw97}
{Navarro} J.~F., {Frenk} C.~S., {White} S.~D.~M., 1997, \apj, 490, 493

\bibitem[{{Robertson} {et~al.}(2006){Robertson}, {Hernquist}, {Cox}, {Di
  Matteo}, {Hopkins}, {Martini}, \& {Springel}}]{robertsonetal06}
{Robertson} B., {Hernquist} L., {Cox} T.~J., {Di Matteo} T., {Hopkins} P.~F.,
  {Martini} P., {Springel} V., 2006, \apj, 641, 90

\bibitem[{{Ross} {et~al.}(2009){Ross}, {Shen}, {Strauss}, {Vanden Berk},
  {Connolly}, {Richards}, {Schneider}, {Weinberg}, {Hall}, {Bahcall}, \&
  {Brunner}}]{rossetal09}
{Ross} N.~P., {Shen} Y., {Strauss} M.~A., {Vanden Berk} D.~E., {Connolly}
  A.~J., {Richards} G.~T., {Schneider} D.~P., {Weinberg} D.~H., {Hall} P.~B.,
  {Bahcall} N.~A., {Brunner} R.~J., 2009, \apj, 697, 1634

\bibitem[{{Salucci} {et~al.}(1999){Salucci}, {Szuszkiewicz}, {Monaco}, \&
  {Danese}}]{saluccietal99}
{Salucci} P., {Szuszkiewicz} E., {Monaco} P., {Danese} L., 1999, \mnras, 307,
  637

\bibitem[{{Seljak}(2000)}]{seljak00}
{Seljak} U., 2000, \mnras, 318, 203

\bibitem[{{Shankar} {et~al.}(2010{\natexlab{a}}){Shankar}, {Crocce},
  {Miralda-Escud{\'e}}, {Fosalba}, \& {Weinberg}}]{shankaretal10}
{Shankar} F., {Crocce} M., {Miralda-Escud{\'e}} J., {Fosalba} P., {Weinberg}
  D.~H., 2010{\natexlab{a}}, \apj, 718, 231

\bibitem[{{Shankar} {et~al.}(2004){Shankar}, {Salucci}, {Granato}, {De Zotti},
  \& {Danese}}]{shankaretal04}
{Shankar} F., {Salucci} P., {Granato} G.~L., {De Zotti} G., {Danese} L., 2004,
  \mnras, 354, 1020

\bibitem[{{Shankar} {et~al.}(2010{\natexlab{b}}){Shankar}, {Weinberg}, \&
  {Shen}}]{shankaretal10b}
{Shankar} F., {Weinberg} D.~H., {Shen} Y., 2010{\natexlab{b}}, \mnras, 406,
  1959

\bibitem[{{Shen} {et~al.}(2007){Shen}, {Strauss}, {Oguri}, {Hennawi}, {Fan},
  {Richards}, {Hall}, {Gunn}, {Schneider}, {Szalay}, {Thakar}, {Vanden Berk},
  {Anderson}, {Bahcall}, {Connolly}, \& {Knapp}}]{shenetal07}
{Shen} Y., {Strauss} M.~A., {Oguri} M., {Hennawi} J.~F., {Fan} X., {Richards}
  G.~T., {Hall} P.~B., {Gunn} J.~E., {Schneider} D.~P., {Szalay} A.~S.,
  {Thakar} A.~R., {Vanden Berk} D.~E., {Anderson} S.~F., {Bahcall} N.~A.,
  {Connolly} A.~J., {Knapp} G.~R., 2007, \aj, 133, 2222

\bibitem[{{Shen} {et~al.}(2009){Shen}, {Strauss}, {Ross}, {Hall}, {Lin},
  {Richards}, {Schneider}, {Weinberg}, {Connolly}, {Fan}, {Hennawi}, {Shankar},
  {Vanden Berk}, {Bahcall}, \& {Brunner}}]{shenetal09}
{Shen} Y., {Strauss} M.~A., {Ross} N.~P., {Hall} P.~B., {Lin} Y., {Richards}
  G.~T., {Schneider} D.~P., {Weinberg} D.~H., {Connolly} A.~J., {Fan} X.,
  {Hennawi} J.~F., {Shankar} F., {Vanden Berk} D.~E., {Bahcall} N.~A.,
  {Brunner} R.~J., 2009, \apj, 697, 1656

\bibitem[{{Sijacki} {et~al.}(2007){Sijacki}, {Springel}, {Di Matteo}, \&
  {Hernquist}}]{sijackietal07}
{Sijacki} D., {Springel} V., {Di Matteo} T., {Hernquist} L., 2007, \mnras, 380,
  877

\bibitem[{{Silk} \& {Rees}(1998)}]{s&r98}
{Silk} J., {Rees} M.~J., 1998, \aap, 331, L1

\bibitem[{{Soltan}(1982)}]{soltan82}
{Soltan} A., 1982, \mnras, 200, 115

\bibitem[{{Spergel} {et~al.}(2003){Spergel}, {Verde}, {Peiris}, {Komatsu},
  {Nolta}, {Bennett}, {Halpern}, {Hinshaw}, {Jarosik}, {Kogut}, {Limon},
  {Meyer}, {Page}, {Tucker}, {Weiland}, {Wollack}, \& {Wright}}]{spergeletal03}
{Spergel} D.~N., {Verde} L., {Peiris} H.~V., {Komatsu} E., {Nolta} M.~R.,
  {Bennett} C.~L., {Halpern} M., {Hinshaw} G., {Jarosik} N., {Kogut} A.,
  {Limon} M., {Meyer} S.~S., {Page} L., {Tucker} G.~S., {Weiland} J.~L.,
  {Wollack} E., {Wright} E.~L., 2003, \apjs, 148, 175

\bibitem[{{Springel}(2005)}]{springel05}
{Springel} V., 2005, \mnras, 364, 1105

\bibitem[{{Springel} \& {Hernquist}(2003)}]{s&h03}
{Springel} V., {Hernquist} L., 2003, \mnras, 339, 312

\bibitem[{{Starikova} {et~al.}(2010){Starikova}, {Cool}, {Eisenstein},
  {Forman}, {Jones}, {Hickox}, {Kenter}, {Kochanek}, {Kravtsov}, {Murray}, \&
  {Vikhlinin}}]{starikovaetal10}
{Starikova} S., {Cool} R., {Eisenstein} D., {Forman} W., {Jones} C., {Hickox}
  R., {Kenter} A., {Kochanek} C., {Kravtsov} A., {Murray} S.~S., {Vikhlinin}
  A., 2010, ArXiv e-prints

\bibitem[{{Teyssier} {et~al.}(2011){Teyssier}, {Moore}, {Martizzi}, {Dubois},
  \& {Mayer}}]{teyssieretal11}
{Teyssier} R., {Moore} B., {Martizzi} D., {Dubois} Y., {Mayer} L., 2011,
  \mnras, 618

\bibitem[{{Thacker} {et~al.}(2009){Thacker}, {Scannapieco}, {Couchman}, \&
  {Richardson}}]{thackeretal09}
{Thacker} R.~J., {Scannapieco} E., {Couchman} H.~M.~P., {Richardson} M., 2009,
  \apj, 693, 552

\bibitem[{{Tinker} {et~al.}(2005){Tinker}, {Weinberg}, {Zheng}, \&
  {Zehavi}}]{tinkeretal05}
{Tinker} J.~L., {Weinberg} D.~H., {Zheng} Z., {Zehavi} I., 2005, \apj, 631, 41

\bibitem[{{Tremaine} {et~al.}(2002){Tremaine}, {Gebhardt}, {Bender}, {Bower},
  {Dressler}, {Faber}, {Filippenko}, {Green}, {Grillmair}, {Ho}, {Kormendy},
  {Lauer}, {Magorrian}, {Pinkney}, \& {Richstone}}]{tremaineetal02}
{Tremaine} S., {Gebhardt} K., {Bender} R., {Bower} G., {Dressler} A., {Faber}
  S.~M., {Filippenko} A.~V., {Green} R., {Grillmair} C., {Ho} L.~C., {Kormendy}
  J., {Lauer} T.~R., {Magorrian} J., {Pinkney} J., {Richstone} D., 2002, \apj,
  574, 740

\bibitem[{{Ueda} {et~al.}(2003){Ueda}, {Akiyama}, {Ohta}, \&
  {Miyaji}}]{uedaetal03}
{Ueda} Y., {Akiyama} M., {Ohta} K., {Miyaji} T., 2003, \apj, 598, 886

\bibitem[{{van den Bosch} {et~al.}(2003){van den Bosch}, {Yang}, \&
  {Mo}}]{vandenboschetal03}
{van den Bosch} F.~C., {Yang} X., {Mo} H.~J., 2003, \mnras, 340, 771

\bibitem[{{Wyithe} \& {Loeb}(2003)}]{w&l03}
{Wyithe} J.~S.~B., {Loeb} A., 2003, \apj, 595, 614

\bibitem[{{Yan} {et~al.}(2004){Yan}, {White}, \& {Coil}}]{yanetal04}
{Yan} R., {White} M., {Coil} A.~L., 2004, \apj, 607, 739

\bibitem[{{Yang} {et~al.}(2003){Yang}, {Mo}, \& {van den Bosch}}]{yangetal03}
{Yang} X., {Mo} H.~J., {van den Bosch} F.~C., 2003, \mnras, 339, 1057

\bibitem[{{Zehavi} {et~al.}(2005){Zehavi}, {Zheng}, {Weinberg}, {Frieman},
  {Berlind}, {Blanton}, {Scoccimarro}, {Sheth}, {Strauss}, {Kayo}, {Suto},
  {Fukugita}, {Nakamura}, {Bahcall}, {Brinkmann}, {Gunn}, {Hennessy},
  {Ivezi{\'c}}, {Knapp}, {Loveday}, {Meiksin}, {Schlegel}, {Schneider},
  {Szapudi}, {Tegmark}, {Vogeley}, \& {York}}]{zehavietal05}
{Zehavi} I., {Zheng} Z., {Weinberg} D.~H., {Frieman} J.~A., {Berlind} A.~A.,
  {Blanton} M.~R., {Scoccimarro} R., {Sheth} R.~K., {Strauss} M.~A., {Kayo} I.,
  {Suto} Y., {Fukugita} M., {Nakamura} O., {Bahcall} N.~A., {Brinkmann} J.,
  {Gunn} J.~E., {Hennessy} G.~S., {Ivezi{\'c}} {\v Z}., {Knapp} G.~R.,
  {Loveday} J., {Meiksin} A., {Schlegel} D.~J., {Schneider} D.~P., {Szapudi}
  I., {Tegmark} M., {Vogeley} M.~S., {York} D.~G., 2005, \apj, 630, 1

\bibitem[{{Zheng} {et~al.}(2005){Zheng}, {Berlind}, {Weinberg}, {Benson},
  {Baugh}, {Cole}, {Dav{\'e}}, {Frenk}, {Katz}, \& {Lacey}}]{zhengetal05}
{Zheng} Z., {Berlind} A.~A., {Weinberg} D.~H., {Benson} A.~J., {Baugh} C.~M.,
  {Cole} S., {Dav{\'e}} R., {Frenk} C.~S., {Katz} N., {Lacey} C.~G., 2005,
  \apj, 633, 791

\bibitem[{{Zheng} {et~al.}(2007){Zheng}, {Coil}, \& {Zehavi}}]{zhengetal07}
{Zheng} Z., {Coil} A.~L., {Zehavi} I., 2007, \apj, 667, 760

\bibitem[{{Zheng} \& {Weinberg}(2007)}]{z&w07}
{Zheng} Z., {Weinberg} D.~H., 2007, \apj, 659, 1

\end{thebibliography}
\bibliographystyle{mn}

\appendix
\end{document}